\documentclass[twocolumn]{aastex631}
\usepackage{amsmath}
\usepackage{booktabs}
\usepackage[version=4]{mhchem} 
\usepackage{gensymb}
\usepackage[T1]{fontenc}        
\usepackage{floatrow}           
\usepackage{tablefootnote}
\usepackage{threeparttable}
\let\tablenum\relax 
\usepackage{siunitx}

\shorttitle{TESS Palomar Survey}
\shortauthors{Gomez Barrientos et al.}

\begin{document}

\title{Validation of TESS Planet Candidates with Multi-Color Transit Photometry and \texttt{TRICERATOPS+}}

\correspondingauthor{Jonathan Gomez Barrientos}
\email{jgomezba@caltech.edu}

\author[0000-0002-0672-9658]{Jonathan Gomez Barrientos}
\affiliation{Division of Geological and Planetary Sciences, California Institute of Technology, Pasadena, CA 91125, USA}

\author[0000-0002-0371-1647]{Michael Greklek-McKeon}
\affiliation{Division of Geological and Planetary Sciences, California Institute of Technology, Pasadena, CA 91125, USA}

\author[0000-0002-5375-4725]{Heather A. Knutson}
\affiliation{Division of Geological and Planetary Sciences, California Institute of Technology, Pasadena, CA 91125, USA}

\author[0000-0002-8965-3969]{Steven Giacalone}
\affiliation{Department of Astronomy, California Institute of Technology, Pasadena, CA 91125, USA}

\author[0000-0002-1422-4430]{W. Garrett Levine}
\affiliation{Department of Astronomy, Yale University. New Haven, CT 06511, USA}
\affiliation{Department of Earth, Planetary, and Space Sciences, University of California, Los Angeles, CA 90095, USA}

\author[0000-0001-9518-9691]{Morgan Saidel}
\affiliation{Division of Geological and Planetary Sciences, California Institute of Technology, Pasadena, CA 91125, USA}

\author[0000-0003-2527-1475]{Shreyas Vissapragada}
\affiliation{Carnegie Science Observatories, 813 Santa Barbara Street, Pasadena, CA 91101, USA}

\author[0000-0002-5741-3047]{David R. Ciardi}
\affiliation{NASA Exoplanet Science Institute, IPAC, California Institute of Technology, Pasadena, CA 91125, USA}

\author[0000-0001-6588-9574]{Karen A.\ Collins}
\affiliation{Center for Astrophysics ${\rm \mid}$ Harvard {\rm \&} Smithsonian, 60 Garden Street, Cambridge, MA 02138, USA}

\author[0000-0001-9911-7388]{David W. Latham}
\affiliation{Center for Astrophysics ${\rm \mid}$ Harvard {\rm \&} Smithsonian, 60 Garden Street, Cambridge, MA 02138, USA}

\author[0000-0001-8621-6731]{Cristilyn N.\ Watkins}
\affiliation{Center for Astrophysics ${\rm \mid}$ Harvard {\rm \&} Smithsonian, 60 Garden Street, Cambridge, MA 02138, USA}

\author[0009-0002-0733-572X]{Polina A. Budnikova}
\affiliation{Sternberg Astronomical Institute, M.V. Lomonosov Moscow State University, 13, Universitetskij pr., 119234, Moscow, Russia}

\author[0009-0003-4203-9667]{Dmitry V. Cheryasov}
\affiliation{Sternberg Astronomical Institute, M.V. Lomonosov Moscow State University, 13, Universitetskij pr., 119234, Moscow, Russia}

\author[0000-0002-4909-5763]{Akihiko Fukui}
\affiliation{Komaba Institute for Science, The University of Tokyo, 3-8-1 Komaba, Meguro, Tokyo 153-8902, Japan}
\affiliation{Instituto de Astrofísica de Canarias (IAC), 38205 La Laguna, Tenerife, Spain}

\author[0000-0001-6637-5401]{Allyson Bieryla}
\affiliation{Center for Astrophysics ${\rm \mid}$ Harvard {\rm \&} Smithsonian, 60 Garden Street, Cambridge, MA 02138, USA}

\author[0000-0002-1836-3120]{Avi Shporer}
\affiliation{Department of Physics and Kavli Institute for Astrophysics and Space Research, Massachusetts Institute of Technology, Cambridge, MA 02139, USA}

\author[0000-0003-2053-0749]{Benjamin M. Tofflemire}
\affiliation{SETI Institute, Mountain View, CA 94043 USA/NASA Ames Research Center, Moffett Field, CA 94035 USA}

\author[0000-0002-2361-5812]{Catherine A. Clark}
\affiliation{NASA Exoplanet Science Institute, IPAC, California Institute of Technology, Pasadena, CA 91125, USA}

\author[0000-0003-2163-1437]{Chris Stockdale}
\affiliation{Hazelwood Observatory, Australia}

\author{Colin Littlefield}
\affiliation{Bay Area Environmental Research Institute, Moffett Field, CA 94035, USA}
\affiliation{NASA Ames Research Center, Moffett Field, CA 94035, USA}

\author[0000-0002-0388-8004]{Emily Gilbert}
\affiliation{Jet Propulsion Laboratory, California Institute of Technology, 4800 Oak Grove Drive, Pasadena, CA 91109, USA}

\author[0000-0003-0987-1593]{Enric Palle}
\affiliation{Instituto de Astrofísica de Canarias (IAC), 38205 La Laguna, Tenerife, Spain}
\affiliation{Departamento de Astrofísica, Universidad de La Laguna (ULL), 38206 La Laguna, Tenerife, Spain}

\author[0000-0002-5443-3640]{Eric Girardin}
\affiliation{Grand Pra Observatory, 1984 Les Hauderes, Switzerland}

\author[0000-0001-9087-1245]{Felipe Murgas}
\affiliation{Instituto de Astrofísica de Canarias (IAC), 38205 La Laguna, Tenerife, Spain}
\affiliation{Departamento de Astrofísica, Universidad de La Laguna (ULL), 38206 La Laguna, Tenerife, Spain}

\author[0000-0003-4500-8850]{Galen J. Bergsten} 
\affiliation{Lunar and Planetary Laboratory, The University of Arizona, Tucson, AZ 85721 USA}

\author[0000-0002-4047-4724]{Hugh~P.~Osborn}
\affiliation{NCCR/Planet-S, Physikalisches Institut, Universität Bern, Gesellschaftsstrasse 6, 3012 Bern, Switzerland}
\affiliation{Department of Physics, ETH Zurich, Wolfgang-Pauli-Strasse 2, CH-8093 Zurich, Switzerland}

\author{Ian J. M. Crossfield}
\affiliation{Department of Physics and Astronomy, University of Kansas, Lawrence, KS 66045, USA}

\author[0000-0002-6424-3410]{Jerome de Leon}
\affiliation{Department of Multi-Disciplinary Sciences, Graduate School of Arts and Sciences, The University of Tokyo, 3-8-1 Komaba, Meguro, Tokyo 153-8902, Japan}

\author[0000-0002-3985-8528]{Jesus Higuera}
\affiliation{U.S. National Science Foundation National Optical-Infrared Astronomy Research Laboratory, 950 N. Cherry Ave., Tucson, AZ 85719, USA}

\author[0000-0002-6480-3799]{Keisuke Isogai}
\affiliation{Okayama Observatory, Kyoto University, 3037-5 Honjo, Kamogatacho, Asakuchi, Okayama 719-0232, Japan}
\affiliation{Department of Multi-Disciplinary Sciences, Graduate School of Arts and Sciences, The University of Tokyo, 3-8-1 Komaba, Meguro, Tokyo 153-8902, Japan}

\author[0000-0002-0885-7215]{Mark~E.~Everett}
\affiliation{U.S. National Science Foundation National Optical-Infrared Astronomy Research Laboratory, 950 N. Cherry Ave., Tucson, AZ 85719, USA}

\author[0000-0003-2527-1598]{Michael B. Lund} 
\affiliation{NASA Exoplanet Science Institute, IPAC, California Institute of Technology, Pasadena, CA 91125, USA}

\author[0000-0001-8511-2981]{Norio Narita}
\affiliation{Komaba Institute for Science, The University of Tokyo, 3-8-1 Komaba, Meguro, Tokyo 153-8902, Japan}
\affiliation{Astrobiology Center, 2-21-1 Osawa, Mitaka, Tokyo 181-8588, Japan}
\affiliation{Instituto de Astrofisica de Canarias (IAC), 38205 La Laguna, Tenerife, Spain}

\author[0000-0001-8227-1020]{Richard P. Schwarz}
\affiliation{Center for Astrophysics ${\rm \mid}$ Harvard {\rm \&} Smithsonian, 60 Garden Street, Cambridge, MA 02138, USA}

\author{Roberto Zambelli} 
\affiliation{Società Astronomica Lunae, Castelnuovo Magra, Italy}

\author[0000-0002-2532-2853]{Steve B. Howell}
\affiliation{NASA Ames Research Center, Moffett Field, CA 94035, USA}

\begin{abstract}

We present an upgraded version of \texttt{TRICERATOPS}, a software package designed to calculate false positive probabilities for planet candidates identified by the Transiting Exoplanet Survey Satellite (TESS). This enhanced framework now incorporates ground-based light curves in separate bandpasses, which are routinely obtained as part of the candidate vetting process. We apply this upgraded framework to explore the planetary nature of 14 TESS planet candidates, combining primarily $J$ band light curves acquired with the 200-inch Hale Telescope at Palomar Observatory with complementary archival observations from the Las Cumbres Observatory Global Telescope (LCOGT), the Fred Lawrence Whipple Observatory (FLWO), and the Teide Observatory, along with existing TESS data and contrast curves from high-resolution imaging. As a result of this analysis we statistically validate (False Positive Probability <$1.5\%$ and Nearby False Positive Probability <$0.1\%$ ) six new planets in five systems: TOI-1346 b, TOI-1346 c, TOI-2719 b, TOI-4155 b, TOI-6000 b, and TOI-6324 b. For these systems, we provide updated estimates of their stellar and planetary properties derived from the TESS and ground-based observations. These new systems contain planets with radii between $0.9-6$ $R_\earth$ and orbital periods between $0.3-5.5$ days. Finally, we use our upgraded version of \texttt{TRICERATOPS} to quantify the relative importance of multi-wavelength transit photometry and high-resolution imaging for exoplanet candidate validation, and discuss which kinds of candidates typically benefit the most from ground-based multi-color transit observations.

\end{abstract}

 
\section{Introduction}
\label{sec:intro}

In recent decades, space-based transit surveys have revolutionized exoplanetary science, vastly expanding our catalog of worlds beyond the solar system and transforming our understanding of exoplanet demographics. Two missions have played a particularly important role in these advances: the Kepler/K2 mission \citep[2009-2018;][]{Borucki2010, Howell2014} and the Transiting Exoplanet Survey Satellite (TESS, 2018-present; \citealt{Ricker2015}).

The primary goal of the Kepler mission was to discover Earth-sized planets in or near the habitable zone of Sun-like stars using the transit method. As a result, it observed a single 100 square degree patch of sky for a period of four years during its primary mission. Over its primary mission, Kepler identified 4,717 planet candidates, 2,781 of which were subsequently confirmed as bona fide planets\footnote{\url{https://exoplanetarchive.ipac.caltech.edu/docs/counts_detail.html}}, unveiling key trends in planet properties that extend to orbital periods of several years. Notably, Kepler's dataset revealed that small planets are abundant around small stars \citep[e.g.,][]{Dressing2015}, that multi-planet systems are common \citep[e.g.,][]{Latham2011,Lissauer2011,Rowe2014, Millholland2017}, and that the radius distribution of the small exoplanet population is bimodal, with two peaks separated by a valley between 1.5 and 2 $R_{\bigoplus}$ \citep[e.g.,][]{Fulton2017, VanEylen2018, Fulton2018}. However, despite these significant advances, our ability to follow up on these discoveries is limited by the fact that many Kepler systems are too faint ($V$ > 14) for radial velocity mass measurements or atmospheric characterization studies.

TESS \citep{Ricker2015} was designed to build on Kepler's success and address its limitations by carrying out an all-sky transit survey that included many more bright stars and a much larger sample of low-mass stars. TESS has identified more than 7,000 planet candidates to date, 622 of which have been confirmed as bona fide planets. This number continues to grow as TESS collects new data during its extended mission phase. The brightest and closest of these TESS systems are excellent targets for observations with the James Webb Space Telescope \citep[JWST; e.g.,][]{Giacalone2022, Mistry2023, Hord2023}.

In order to carry out demographic studies on the population of transiting planets identified by these surveys, it is crucial to first rule out as many astrophysical false positives as possible in the sample of transiting planet candidates. These typically include a mix of eclipsing binaries and blended eclipsing binaries. Vetting approaches have evolved to operate at both the pixel and light curve level using the original survey photometry alone. Pixel-level analysis, such as the centroid test \citep[e.g.,][]{Batalha2010, Bryson2013, Gunther2017}, searches for offsets in the host star's centroid during a transit, which can reveal blends involving background stellar eclipsing binaries. At the light curve level, the odd-even test \citep[e.g.,][]{McCauliff2015, Coughlin2016, Kostov2019, Zink2020} compares transit depths between alternating events to identify eclipsing binaries masquerading as planets with twice their true orbital period.

Although more expensive to carry out, ground-based observations also play a central role in the vetting of transiting planet candidates. High-resolution imaging \citep[e.g.,][]{Law2014, Baranec2016, Furlan2017, Hirsch2017, Ziegler2017, Ziegler2018, Ziegler2020, Ziegler2021, Howell2021, Lester2021} can identify stellar companions that may be diluting the transit signal or are instead the source of the transit signal within fractions of an arcsecond of the target star. High-resolution spectroscopy can also identify systems with blended lines indicative of multiple stellar components \citep[e.g.,][]{Kolbl2015}. Since transiting planet light curves are largely achromatic, ground-based observations in different broadband filters can also rule out or identify false positives by measuring the wavelength-dependent transit depth \citep[e.g.,][]{Drake2003, Colon2012, Desert2015, Parviainen2019, Pelaez-Torres2024, Lillo-Box2024}. We can combine all of the available information about a system in order to calculate the probability that a given transiting planet candidate might be a false positive. This approach is known as `statistical validation' and has been applied to large samples of candidates from the Kepler, K2, and TESS missions \citep[e.g.,][]{Rowe2014, Thompson2018, Zink2020, Zink2021, Kunimoto2020, Kunimoto2024}. For example, statistical tools developed during the Kepler era such as VESPA \citep{Morton2012, Morton2016}, PASTIS \citep{Diaz2014, Santerne2015}, and BLENDER \citep{Torres2004, Torres2005, Torres2011} are able to incorporate results from follow-up observations (e.g., high-resolution imaging, spectroscopy, and multi-color transit photometry). More recently, these tools have been replaced by tools with lower computation times such as \texttt{TRICERATOPS} \citep{Giacalone2021}, which can also incorporate results from high-resolution imaging. However, this package currently lacks an option to incorporate multi-color transit observations.

In this study, we upgrade \texttt{TRICERATOPS} to incorporate multi-color transit photometry in calculating false positive probabilities for planet candidates. We utilize this enhanced framework to explore the planetary nature of 14 TESS planet candidates, primarily using infrared light curves obtained with the Wide Field InfraRed Camera (WIRC; \citealt{Wilson2003}) on the 200-inch Hale Telescope at Palomar Observatory. We complement these observations with data from the Las Cumbres Observatory Global Telescope (LCOGT), the Fred Lawrence Whipple Observatory (FLWO), and the Teide Observatory. We begin in Section \ref{sec:data} by describing our observations. We proceed to describe our light curve analysis in Section \ref{sec:fitting} and modifications to \texttt{TRICERATOPS} in Section \ref{sec:tricer}. Lastly, we present our results in Sections \ref{sec:results} and \ref{sec:results2}, discuss their implications in Section \ref{sec:discussion}, and summarize our conclusions in Section \ref{sec:conclusion}.

\section{Observations}
\label{sec:data}

\begin{table*}[]
\centering
\fontsize{8pt}{8pt}\selectfont
\begin{tabular}{llcccccccccc}
\hline
TOI & TIC ID & RA & Dec & $TESS$ & $V$ & $J$ & $T_\mathrm{eff}$ & $\log g$ & [m/H] & $R_*$ & $M_*$ \\ 
    & & & & (mag) & (mag) & (mag) & (K) & & (dex) & ($R_\odot$) & ($M_\odot$) \\
\hline
1254$^a$ & 236714379 & 15:56:48.35 & 65:53:19.82 & 10.79 & 11.65 & 10.04 & $5666 \pm 50$ & $4.54 \pm 0.10$ & $0.15 \pm 0.08$ & $1.09 \pm 0.06$ & $0.95 \pm 0.12$ \\
1346$^a$ & 219852882 &17:06:29.44 & 68:50:35.89 & 10.73 & 11.69 & 9.91 & $5099 \pm 50$ & $4.61 \pm 0.10$ & $-0.01 \pm 0.08$ & $0.78 \pm 0.05$ & $0.82 \pm 0.10$ \\
1616$^a$ & 322054600 &21:35:28.84 & 68:59:14.31 & 10.45 & 10.83 & 10.02 & $6505 \pm 72$ & $4.25 \pm 0.12$ & $0.34 \pm 0.08$ & $1.45 \pm 0.06$ & $1.43 \pm 0.26$ \\
2719$^a$ & 176314383 &04:46:13.19 & -00:48:17.54 & 12.07 & 12.68 & 11.41 & $5762 \pm 50$ & $4.22 \pm 0.10$ & $0.27 \pm 0.08$ & $1.69 \pm 0.10$ & $1.02 \pm 0.13$ \\
4051 & 237101326 & 16:02:46.04 & 71:13:27.44 & 13.03 & 13.81 & 12.21 & $5091 \pm 122$ & $4.36$ & - & $1.00$ & $0.85$ \\
4094$^a$ & 280035202 & 19:50:42.76 & 68:10:03.58 & 11.10 & 11.21 & 10.55 & $5998 \pm 50$ & $4.27 \pm 0.10$ & $0.21 \pm 0.08$ & $1.24 \pm 0.06$ & $1.18 \pm 0.23$ \\
4155$^a$ & 467331291 & 21:42:32.69 & 77:44:03.78 & 11.25 & 11.85 & 10.61 & $5724 \pm 50$ & $4.44 \pm 0.10$ & $0.11 \pm 0.08$ & $0.95 \pm 0.05$ & $1.16 \pm 0.22$ \\
4731$^a$ & 438368981 & 06:28:35.34 & 14:56:25.39 & 11.80 & 12.40 & 11.32 & $6014 \pm 75$ & $4.36 \pm 0.13$ & $0.24 \pm 0.08$ & $1.25 \pm 0.17$ & 1.36  \\
5706$^a$ & 137843225 & 15:06:56.46 & 57:29:44.49 & 10.76 & 11.64 & 9.81 & $4673 \pm 50$ & $4.69 \pm 0.10$ & $-0.28 \pm 0.08$ & $1.05 \pm 0.08$ & $0.69 \pm 0.08$ \\
5735 & 157769780 & 09:48:30.01 & 77:23:23.09 & 12.28 & 15.00 & 10.65 & $3222 \pm 157$ & $5.04 \pm 0.02$ & - & $0.21 \pm 0.01$ & $0.18 \pm 0.02$ \\
6000 & 259233660 & 19:30:25.51 & 68:09:16.53 & 13.16 & 15.45 & 11.66 & $3419 \pm 157$ & $4.85 \pm 0.00$ & - & $0.38 \pm 0.01$ & $0.37 \pm 0.02$ \\
6324 & 372207328 & 22:03:22.69 & 67:29:55.24 & 10.97 & 13.39 & 9.41 & $3358 \pm 157$ & $4.93 \pm 0.01$ & - & $0.30 \pm 0.01$ & $0.27 \pm 0.02$ \\
6397 & 359629653 & 18:40:15.56 & 58:59:15.03 & 13.18 & 13.90 & 12.38 & $5041 \pm 122$ & $4.43$ & - & $0.93$ & $0.84$ \\
\hline
\end{tabular}
\caption{Summary of the target stellar properties}
\label{tab:stellar_props}
\vspace{0.1cm}
\footnotesize
\begin{flushleft}
\textbf{Notes.}
\tablenotemark{a} Teff, log(g) and m/H were derived from TRES spectra using the Stellar Parameter Classification (SPC) tool as described in Section \ref{sec:spec}. All other values are from the TESS Input Catalog (TIC; \citealt{Stassun2019}). The targets TOI-4051 and TOI-6397 lack reported uncertainties for $\log g$, $R_*$, and $M_*$ in the TIC catalog. The derivation of these values is detailed in \citealt{Stassun2019} and \citealt{Paegert2021}.
\end{flushleft}
\end{table*}

\subsection{Palomar Data}
\label{sec:palomar}

Over a period of two years, we obtained transit observations of 13 TESS planet candidates in the $J$ band using the Wide-field InfraRed Camera (WIRC, \citealt{Wilson2003}) on the 200-inch Hale Telescope at Palomar Observatory (see Table \ref{tab:palomar_obs} for observation details). These candidates were observed serendipitously during the downtime of nights allocated to other targets, allowing us to maximize the efficiency of our telescope time.

All of our transit observations used a custom beam-shaping diffuser, which produces a top-hat point spread function (PSF) with a full width at half-maximum of 3\arcsec. With the diffuser, we achieved largely stable PSFs throughout the duration of our observations, mitigating the impact of time-correlated noise from changing conditions \citep{Stefansson2017, Vissapragada2020}. The diffuser also enabled us to increase our exposure times for bright targets without saturating the detector, and therefore to achieve a higher overall observing efficiency.

We extracted light curves for each target and a set of comparison stars using \texttt{ExoWIRC}\footnote{\url{https://exowirc.readthedocs.io/en/latest/}}. After flat-fielding, dark subtracting, and correcting for bad pixels as described in \cite{Vissapragada2020}, \texttt{ExoWIRC} uses the Python package \texttt{photutils} \citep{Bradley2020} to perform circular aperture photometry on every star in the image with a signal-to-noise ratio higher than a manually specified threshold. In our sample, this threshold typically ranges from 100 to 500 depending on the target star brightness and field density. To optimize the photometric precision, we used \texttt{ExoWIRC} to test various aperture sizes with radii between $5-30$ pixels. For each aperture size in this range, we extracted the target and comparison star light curves, detrended the target light curve using the average light curve of the comparison stars, and computed the root mean square (rms) of the detrended light curve. We then adopted the aperture that minimized the rms of the target light curve in our subsequent analysis. 

\subsection{TESS Data}
\label{sec:tess}

We downloaded the publicly available TESS observations of 14 targets (see Table \ref{tab:stellar_props}) from the Mikulski Archive for Space Telescopes using the \texttt{lightkurve} package \citep{2018ascl.soft12013L}. These observations undergo vetting by the TESS Science Office through search pipelines before being issued as TOI alerts \citep{Guerrero2021}. In this work, we used all available Presearch Data Conditioning Simple Aperture Photometry (PDCSAP; \citealt{Stumpe2012, Smith2012, Stumpe2014}) light curves at 2-minute cadence that were provided by the TESS Science Processing Operations Center (SPOC) pipeline \citep{Jenkins2016}. We note that 2-minute SPOC light curves are not currently available for TOI-2719.01 as this object was not targeted for a 2-minute postage stamp in the TESS prime mission. Thus, for this target, we instead used the processed light curves at 10-minute cadence from Sector 32, which were provided by the Quick-Look Pipeline \citep[QLP;][]{Huang2020, Kunimoto2022}. We note that processed light curves of TOI-2719.01 at a 30-minute cadence from Sector 5 are also available, but are omitted from our analysis since the 10-minute cadence data provides better temporal resolution.

After downloading the processed TESS observations for each planet candidate, we masked the individual transits and implemented a Lomb–Scargle periodogram \citep{Lomb1976, Scargle1982} to look for residual periodic trends that could affect the quality of the transit signals. Although some 2-minute light curves from Sectors 14-26 contain an uncorrected bias in the background subtraction\footnote{See Section 4.2, page 9 in, \url{https://archive.stsci.edu/missions/tess/doc/tess_drn/tess_sector_27_drn38_v02.pdf}}, we verified that this systematic effect does not impact our analysis. Our periodogram analysis revealed that TOI-12541.01, TOI-1346.01, TOI-1346.02, TOI-4155.01, and TOI-4731.01 exhibit trends. To obtain the most accurate TESS light curves for these candidates, we modeled the trends using Gaussian convolution, as described in Section 3.1 of \citealt{Greklek-McKeon2023}. We then divided by the model trend to flatten the TESS light curves. Finally, for each candidate other than TOI-5735.01, we phase-folded the TESS data on the orbital period, \textit{P}, and the mid-transit time, $t_0$, reported on ExoFOP\footnote{\url{https://exofop.ipac.caltech.edu/tess/}} and truncated the data so that the total coverage is 3$\times$ the transit duration. In the case of TOI-5735.01, we found that phase-folding the TESS light curve using the currently reported values yields an asymmetric transit shape with a low signal-to-noise ratio. As discussed in Section \ref{sec:fitting}, we therefore followed a different approach to modeling the TESS light curve of this candidate.

\subsection{LCOGT Data}

The Las Cumbres Observatory Global Telescope \citep[LCOGT;][]{Brown:2013} 1\,m network nodes used in the observations are located at Teide Observatory on the island of Tenerife (TEID) and McDonald Observatory near Fort Davis, Texas, United States (McD) and the 2\,m network node is at the Faulkes Telescope North at Haleakala Observatory on Maui, Hawai'i. The 1\,m telescopes are equipped with $4096\times4096$ SINISTRO cameras that have an image scale of $0\farcs389$ per pixel, resulting in a $26\arcmin\times26\arcmin$ field of view. The 2\,m telescope is equipped with the MuSCAT3 multi-band imager \citep{Narita:2020}. All LCOGT images were calibrated by the standard LCOGT {\tt BANZAI} pipeline \citep{McCully:2018} and differential photometric data were extracted using {\tt AstroImageJ} \citep{Collins:2017}.

\subsection{KeplerCam}

KeplerCam is installed on the 1.2\,m telescope at the Fred Lawrence Whipple Observatory (FLWO) at Mt. Hopkins, Arizona. The $4096\times4096$ Fairchild CCD 486 detector has an image scale of $0\farcs672$ per $2\times2$ binned pixel, resulting in a $23\farcm1\times23\farcm1$ field of view. Photometric data were extracted using {\tt AstroImageJ}. We used the {\tt TESS Transit Finder}, which is a customized version of the {\tt Tapir} software package \citep{Jensen:2013}, to schedule our transit observations. The light curve data are available on the ExoFOP website$^4$ and are tabulated in Table \ref{tab:lco_keplercam_obs}.

\subsection{MuSCAT2 observations}

TOI-4731 was observed on the night of 14 February 2022 using the multi-band imager MuSCAT2 \citep{Narita2019}, mounted on the 1.5 m Telescopio Carlos S\'{a}nchez (TCS) at Teide Observatory, Spain. MuSCAT2 is equipped with four CCDs, enabling simultaneous imaging in the $g'$, $r'$, $i'$, and $z_s$ bands with minimal readout time. Each CCD has a resolution of 1024 × 1024 pixels, covering a field of view of $7.4 \times 7.4$ arcmin$^2$.

Observations were conducted with the telescope in nominal focus; however, the $i'$-band camera experienced connection issues and was unavailable. Exposure times were set to 20, 7, and 25 seconds for the $g'$, $r'$, and $z_s$ bands, respectively. Some post-transit observations were affected by dome vignetting and were excluded from the transit fit.

The raw data were processed using the MuSCAT2 pipeline \citep{Parviainen2019}, which performs dark and flat-field calibrations, aperture photometry, and transit model fitting while accounting for instrumental systematics. The pipeline determined an optimal photometric aperture of 10.9\arcsec, which includes flux from the nearby star TIC 717538186. To properly deblend the light curves for analysis with \texttt{TRICERATOPS}, we estimated the flux ratio between TOI-4731 and TIC 717538186 in each bandpass. Since the stars are closely spaced, we used PSF fitting to determine their fluxes and computed the flux ratio in each bandpass. For this, we used the \texttt{Photutils} package from \texttt{Astropy} to fit a circular two-dimensional Gaussian PSF model to the science images. The procedure began with background correction using the \texttt{Background2D} routine, which included a 3$\sigma$ sigma-clipping threshold and a median background estimator. The PSF model had three free parameters: the positions of the stars, the total flux, and the full width at half maximum (FWHM) of the PSF. To facilitate convergence, we initialized the star positions using the centroid coordinates from the $r'$ band, where the stars were better resolved. We applied the PSF model fit to each image in the time series and derived the flux ratios between TOI-4731 and TIC 717538186. From the time-series data, we calculated the median and 1$\sigma$ confidence intervals for the measured flux ratios in each bandpass. The results are as follows: $g'_{\mathrm{ratio}} = 1.87^{+0.14}_{-0.23}$, $r'_{\mathrm{ratio}} = 1.83^{+0.07}_{-0.16}$, and $z_{s\;{\mathrm{ratio}}} = 1.72^{+0.08}_{-0.18}$.

\subsection{High-resolution imaging}
\label{sec:high-contrast}

\begin{figure*}
    \centering
    \includegraphics[width=0.8\textwidth]{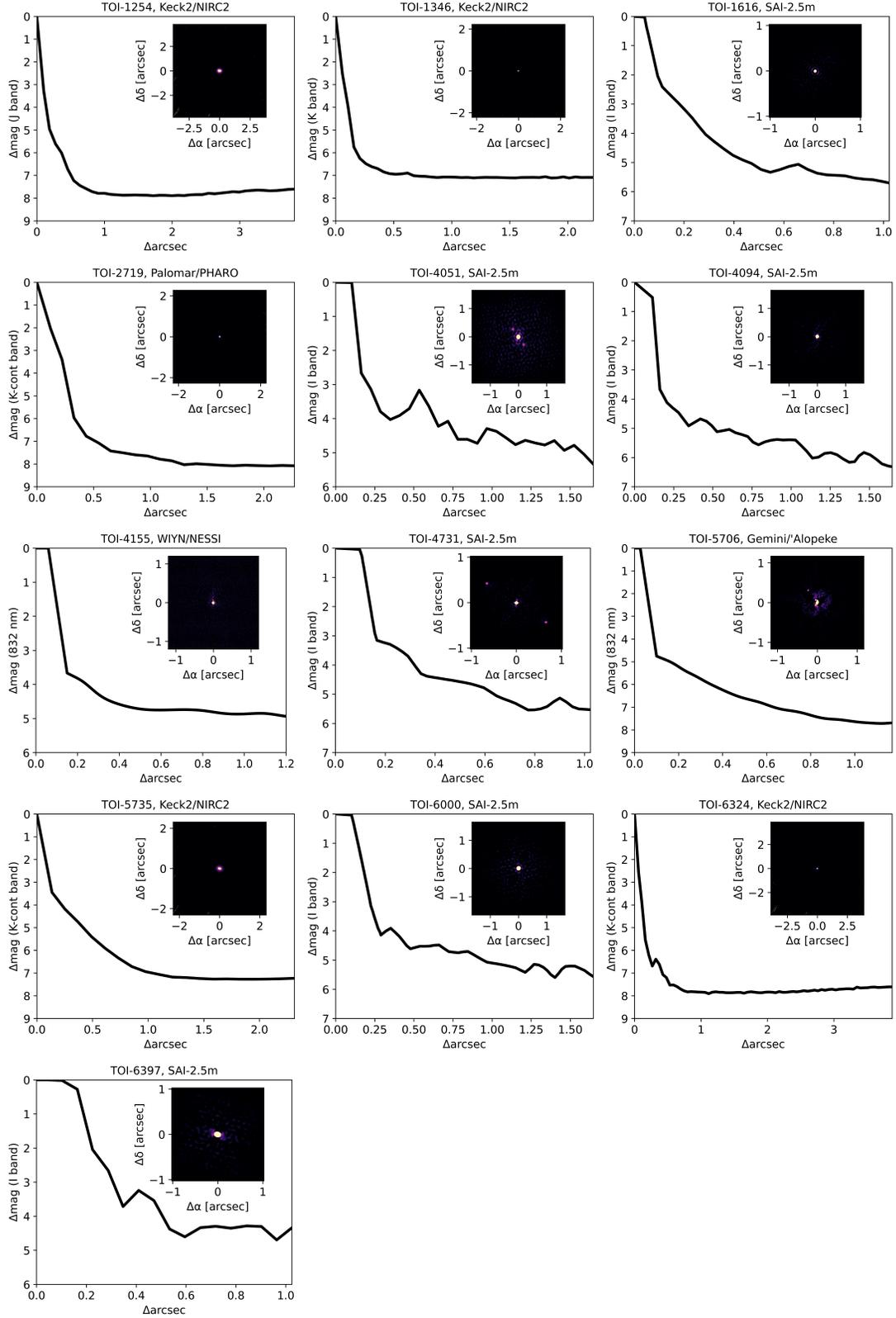}
    \caption{Sensitivity curves (5$\sigma$ contrast) for our targets from high-resolution imaging. The insets show the high resolution images. Companions are seen near TOI-4051, TOI-4731, and TOI-5706.}
    \label{fig:ccs}
\end{figure*}

We downloaded the available contrast curves for each target from the ExoFOP website$^4$. We were able to obtain high-resolution images for all targets, and summarize the properties of these observations in Table \ref{tab:imaging}. These observations revealed that TOI-4051, TOI-4731, and TOI-5706 have nearby stellar companions. The companion to TOI-4051 is 3.1 magnitudes fainter in $I$ band and has a projected separation of 0.53\arcsec. The companion to TOI-4731 is 2.3 magnitudes fainter in $I$ band and has a projected separation of 2.01\arcsec. The companion to TOI-5706 is 3.7 magnitudes fainter in the $K$-continuum filter and is separated by 0.43\arcsec. For our \texttt{TRICERATOPS} analyses, we selected the observation with the best contrast sensitivity for each target. This typically corresponded to the facility with the largest telescope aperture. In some instances, the largest aperture telescope collected a contrast curve in a red and a blue filter. In these cases we found that the red filter provided a higher contrast sensitivity. The selected contrast curves and their corresponding images are shown in Figure \ref{fig:ccs} and highlighted in bold in Table \ref{tab:imaging}.

\subsection{TRES Spectra}
\label{sec:spec}

We obtained high resolution spectrosopy of all of our target stars using the 1.5m Tillinghast Reflector Echelle Spectrograph (TRES, \citealt{tres}) located at the Fred Lawrence Whipple Observatory (FLWO) in Arizona, USA. TRES is a fiber-fed echelle spectrograph with a wavelength range of $390-910$~nm and a resolving power of $R=$44,000. Spectra were extracted as described in \cite{buchhave2010} and then used to derive stellar parameters using the Stellar Parameter classification tool (SPC, \citealt{buchhave2012}). SPC cross-correlates each observed spectrum against a grid of synthetic spectra based on Kurucz atmospheric models \citep{kurucz1992} in order to determine the effective temperature, surface gravity, metallicity, and rotational velocity of the star. The stellar parameters listed in Table \ref{tab:stellar_props} are the average of the stellar parameters calculated from multiple spectra taken at different times.

\section{Light Curve Modeling}
\label{sec:fitting}

\begin{table*}
\centering
\begin{tabular}{lllll}
\hline
TOI & $t_0$ - 2450000$~^a$ & $a/R_*~^b$ & $b$ & $R_p/R_*$ \\
\hline
1254.01 & $\mathcal{N}(9601.9909,0.0003)$ & $\mathcal{N}(3.83,0.20)$ & $\mathcal{U}(0.0,1.2)$ & $\mathcal{U}(0.0,0.2)$ \\
1346.01 & $\mathcal{N}(9766.2249, 0.0023)$ & $\mathcal{N}(15.06, 0.40)$ & $\mathcal{U}(0.0,1.0)$ & $\mathcal{U}(0.0,0.2)$ \\
1346.02 & $\mathcal{N}(9978.0235,0.0030)$ & $\mathcal{N}(7.33,0.46)$ & $\mathcal{U}(0.0,1.0)$ & $\mathcal{U}(0.0,0.1)$ \\
1616.01 & $\mathcal{N}(9821.9649,0.0013)$ & $\mathcal{N}(3.98,0.17)$ & $\mathcal{U}(0.0,1.0)$ & $\mathcal{U}(0.0,0.1)$ \\
2719.01 & $\mathcal{N}(9882.9536,0.0182)$ & $\mathcal{N}(5.63,0.32)$ & $\mathcal{U}(0.0,1.0)$ & $\mathcal{U}(0.0,0.1)$ \\
4051.01 & $\mathcal{N}(9639.0106,0.0047)$ & $\mathcal{N}(5.29,0.53)$ & $\mathcal{U}(0.0,1.0)$ & $\mathcal{U}(0.0,0.2)$ \\
4094.01 & $\mathcal{N}(9897.6699,0.0437)$ & $\mathcal{N}(10.39,0.48)$ & $\mathcal{U}(0.0,1.0)$ & $\mathcal{U}(0.0,0.1)$ \\
4155.01 & $\mathcal{N}(9803.9383,0.0025)$ & $\mathcal{N}(10.39,0.50)$ & $\mathcal{U}(0.0,1.0)$ & $\mathcal{U}(0.0,0.1)$ \\
4731.01 & $\mathcal{N}(9600.7692,0.0003)$ & $\mathcal{N}(4.79,0.65)$ & $\mathcal{U}(0.0,1.2)$ & $\mathcal{U}(0.0,0.2)$ \\
5706.01 & $\mathcal{N}(9822.6503,0.0004)$ & $\mathcal{N}(1.78,0.13)$ & $\mathcal{U}(0.0,1.2)$ & $\mathcal{U}(0.0,0.2)$ \\
5735.01 & $\mathcal{N}(9968.7605,0.0056)$ & $\mathcal{N}(18.60,2.70)$ & $\mathcal{U}(0.0,1.2)$ & $\mathcal{U}(0.0,0.2)$ \\
6000.01 & $\mathcal{N}(10109.9171, 0.0021)$ & $\mathcal{N}(4.67,0.14)$ & $\mathcal{U}(0.0,1.0)$ & $\mathcal{U}(0.0,0.1)$ \\
6324.01 & $\mathcal{N}(10110.9547, 0.0007)$ & $\mathcal{N}(3.90,0.60)$ & $\mathcal{U}(0.0,1.0)$ & $\mathcal{U}(0.0,0.2)$ \\
6397.01 & $\mathcal{N}(10098.6892,0.0005)$ & $\mathcal{N}(6.04,0.60)$ & $\mathcal{U}(0.0,1.2)$ & $\mathcal{U}(0.0,0.3)$ \\
\hline
\end{tabular}
\caption{Priors for the light curve model of each target}
\label{tab:priors}
\vspace{0.05cm}
\footnotesize
\textbf{Notes.} $^a$ Obtained from the ExoFOP website and the TESS Transit Finder Tool \citep{Jensen2013}\\ 
$^b$ Derived from propagating the uncertainties in $R_*$, $M_*$, and $P$ through Kepler's third law. 
For all targets, the jitter prior log($\sigma_\mathrm{systematic}$) = $\mathcal{{U}}({-6},{-2})$ and the TESS error scaling prior $k$ = $\mathcal{{U}}({0.5},{1.5})$. 
\end{table*}

Having collected transit light curves across multiple facilities and wavelengths, we explored whether the transit shapes and depths of our targets are statistically consistent across 0.50 to 1.35 $\micron$ (TESS band: 0.6-1.0 $\micron$; LCOGT $g$, $r$, $i$, $z$ optical bands; and Palomar $J$ band: 1.15-1.35 $\micron$). For most targets, we performed a joint fit of the phase-folded TESS data and the Palomar data, and if available, separately performed a joint fit of the phase-folded TESS data and the LCOGT/FLWO/MUSCAT2 observations. For TOI-5735.01, where the SPOC-ExoFOP period estimate is suboptimal, we fit the individual TESS transits and allow the transit epoch and period to vary as free parameters rather than phase-folding the data with the reported ephemeris.

We used \texttt{ExoWIRC} to model these transit light curves. This package utilizes \texttt{exoplanet} \citep{Foreman-Mackey2021}, which couples the light curve model \texttt{Starry} \citep{Luger2019} to \texttt{PyMC3}'s \citep{Salvatier2015} No U-turn Sampling \citep[][]{Hoffman2011} algorithm to explore the posterior distribution. In all of our fits, we set the orbital eccentricity to zero. Since all of the candidates in our sample are close-in, we expect most to have tidally circularized orbits. We additionally note that small orbital eccentricities ($<0.1$) will have a negligible effect on the shape of the transit light curve at the signal-to-noise ratio of our data \citep[see e.g.,][]{Winn2010}. Our transit model is parameterized by global variables for the semi-major axis in stellar radius units ($a/R_*$), impact parameter (\textit{b}), and the mid-transit time ($t_0$). We account for the uncertainties on the orbital period ($P$) by including it as a free parameter with a Gaussian prior bounded by the uncertainties reported in the TESS ephemeris on ExoFOP$^4$. In addition to these global parameters, our joint fits of the TESS and Palomar data contain two separate planet-star radius ratios ($R_p/R_*$) and error scaling terms. For the TESS data, we fit a parameter $k$ that scales the original TESS error bars to match the scatter in the residuals so that $\sigma_\mathrm{true} = \sigma \cdot k$. For Palomar, where the uncertainties on individual data points can vary depending on the flux within a given integration, we fit a jitter term that is added in quadrature to the photon noise ($\sigma^2 = \sigma_\mathrm{poisson}^2 + \sigma_\mathrm{systematic}^2$). We use a quadratic limb-darkening law with coefficients $u_1$ and $u_2$ calculated for each bandpass using \texttt{ExoTIC-LD} \citep{david_grant_2022_7437681}\footnote{\url{https://github.com/Exo-TiC/ExoTiC-LD}}, adopting the stellar properties of each target from Table \ref{tab:stellar_props}. Given the signal-to-noise ratio of our light curves, we do not expect these coefficients to be well-constrained by the data. As a result, including them as free parameters in our fit could bias our measured transit depths \citep[see e.g.,][]{Coulombe2024}. We list the priors for all the free parameters in our joint fits of the TESS and Palomar data in Table \ref{tab:priors}.

We also included a systematic noise model in our fit to the Palomar data to account for variations in the light curve due to changes in atmospheric conditions (e.g., airmass, telescope tracking, and instrumental effects). This systematic noise model includes a weighted sum of light curves from nearby stars in the field, where the weights are free parameters in the fit. As discussed in \cite{Vissapragada2020} and \cite{Greklek-McKeon2023}, we found that we obtained optimal results for most targets when we detrend with up to ten of the comparison stars whose light curves most closely track that of our target star. In general, we found that the best comparison stars are those with a brightness similar to that of the target star. For some nights of data, we also detrended using linear functions of the airmass, PSF width, background flux, and/or the distance from the median centroid. For each night of data, we selected the combination of decorrelation parameters that minimized the Bayesian Information Criterion (BIC; \citealt{Schwarz1978}). 

For the LCOGT, FLWO/KeplerCam, and MUSCAT2 observations, we employed the same light curve model (that is, $a/R_*$, \textit{b}, $t_0$, $P$, and $R_p/R_*$ are free parameters) and fitting approach as described above (i.e., orbital eccentricity set to zero and fixed limb-darkening). However, rather than fitting all datasets simultaneously for a given target, we performed separate joint-fits between the TESS data and each individual bandpass. Although it would have been preferable to carry out a comprehensive joint-fit incorporating all datasets simultaneously, this would have required a prohibitively large number of free parameters in many cases.  Our goal for the joint fitting was to obtain the best possible constraints on the other transit shape parameters in order to reduce the corresponding uncertainties on the band-specific $R_p/R_*$ values.  In this case, the phased TESS light curve already provided a strong constraint on the transit shape, and there was relatively little added information to be gained by jointly fitting additional ground-based light curves.  We therefore concluded that jointly fitting the TESS light curve with each individual ground-based bandpass would produce equivalent results with fewer free parameters. As with the Palomar+TESS fit described previously, we allowed for separate 
$R_p/R_*$ parameters for TESS and each ground-based dataset.

We report the results of our light curve modeling procedure in Table \ref{tab:posterior}. For each free parameter in the transit model, we list the median value and the 68$\%$ confidence interval from the posterior. We find that TOI-1254.01, TOI-1616.01, TOI-4051.01, and TOI-5706.01 show evidence of chromatic transit depths across the various data sets. Although the transit depths of the remaining candidates appear to be in better agreement, this fact alone is not sufficient to establish that they must be transiting planets. In the following section, we outline our framework for quantifying the probability that a transit signal is a false positive when multi-color photometry is available.

\section{Statistical Validation with Multi-color Photometry}
\label{sec:tricer}

\begin{figure*}
    \centering
    \includegraphics[width=0.99\textwidth]{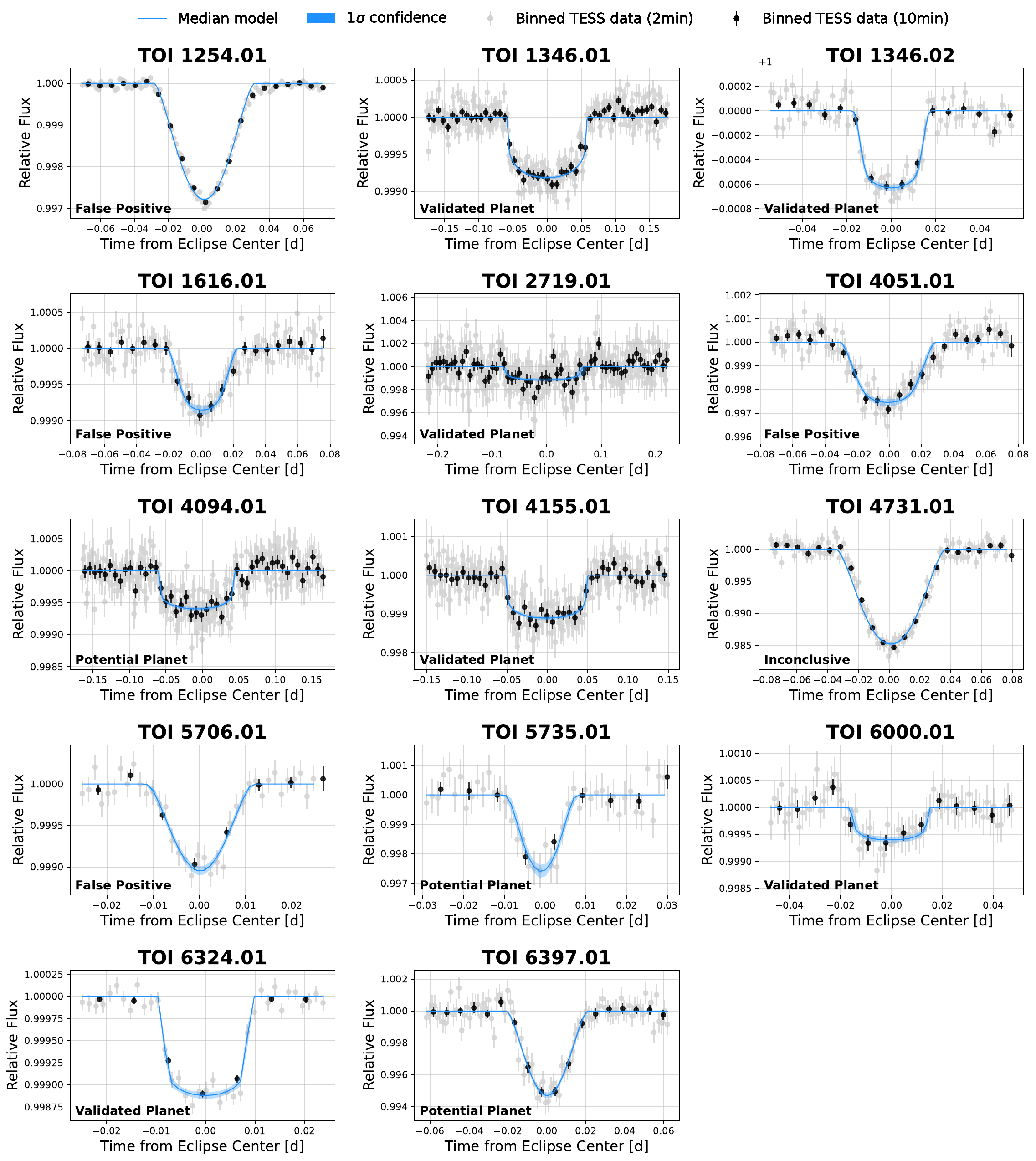}
    \caption{Phase-folded light curves from TESS. The data are binned in 2-minute (gray) and 10-minute (black) intervals. The solid blue line depicts the median posterior model, with light blue shading indicating the 1$\sigma$ (68$\%$) confidence interval.}
    \label{fig:tess_obs}
\end{figure*}

\begin{figure*}
    \centering
    \includegraphics[width=0.99\textwidth]{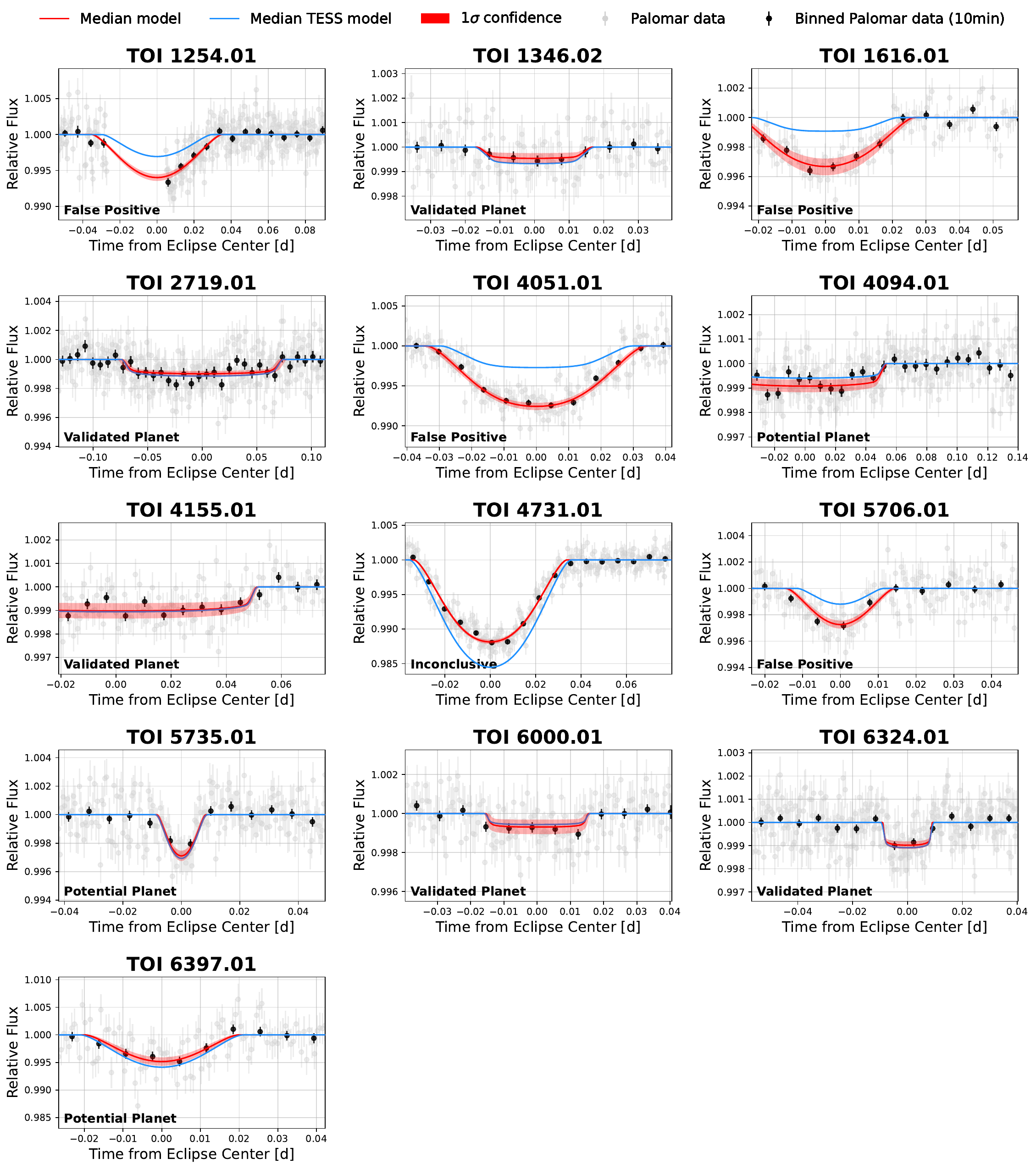}
    \caption{$J$ band light curves from Palomar. The gray data points correspond to the unbinned data and the black data points correspond to data binned in 10-min intervals. The solid red line is the median model from the posterior distribution. The light red shading corresponds to the 1$\sigma$ (68$\%$) confidence interval. The solid blue curve shows the median TESS light curve projected into the $J$ band by applying $J$ band limb darkening coefficients.}
    \label{fig:palomar_obs}
\end{figure*}

\begin{figure*}
    \centering
    \includegraphics[width=1.01\textwidth]{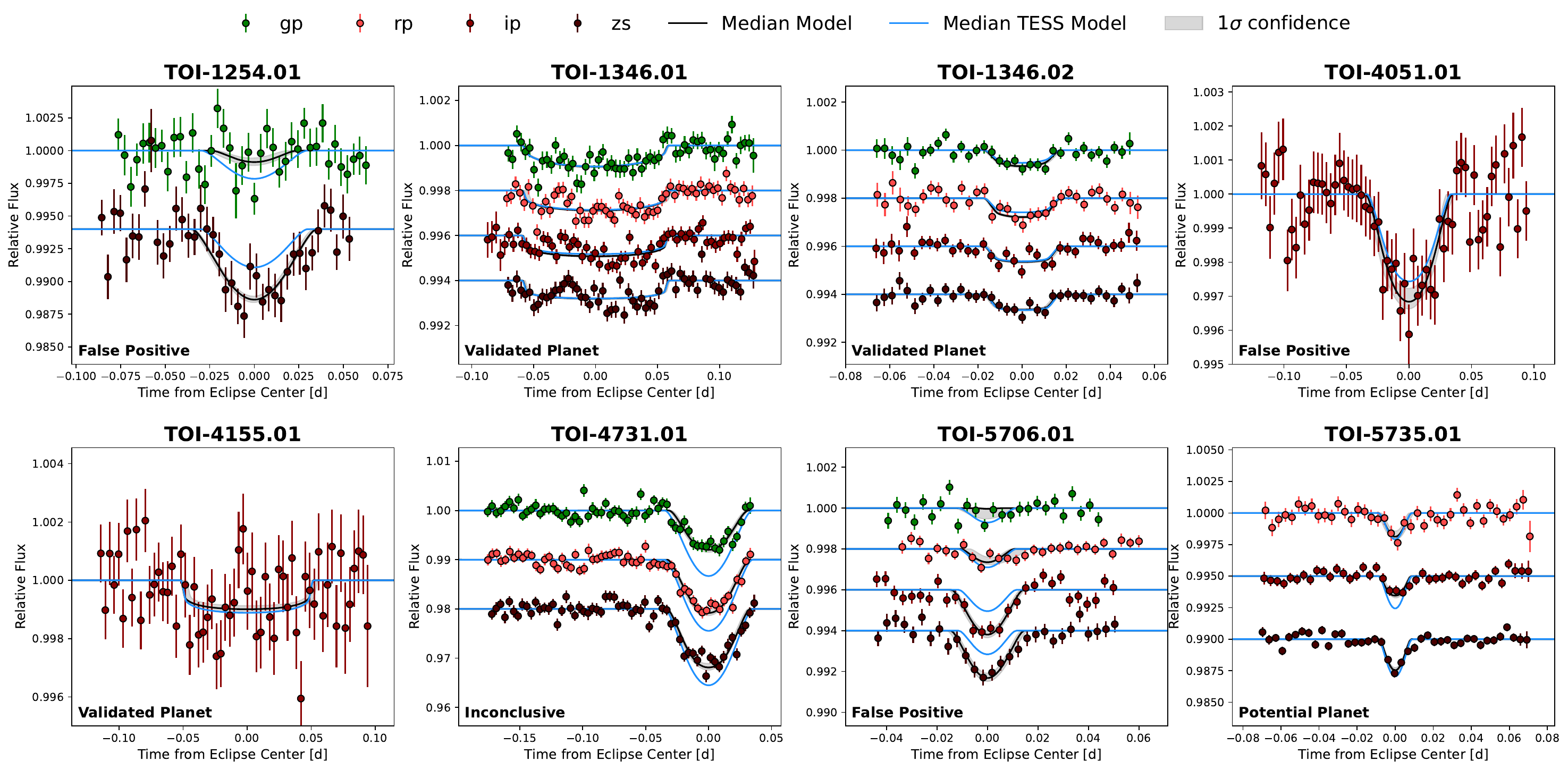}
    \caption{Ground-based light curves from LCOGT, KeplerCam, and MUSCAT2. The colored data points correspond to the binned data at 5 minute cadence. The solid black line is the median model from the posterior distribution. The solid blue curve is the median TESS light curve projected into the corresponding band.}
    \label{fig:lco_obs}
\end{figure*}

\texttt{TRICERATOPS} \citep{Giacalone2021} is a Bayesian code that models the light curves of transiting planets and false positive scenarios. The code incorporates prior information about the population of stars in the Milky Way to compute the false positive probability of a given planet candidate. For instance, \texttt{TRICERATOPS} applies a prior cut to exclude companions with mass ratios $<$ 0.1, which corresponds to companions with M $<$ 0.1$\mathrm{M_\odot}$ for typical GKM hosts. This is because close-in companions with masses below this cutoff are very rare (this is the brown dwarf desert, see \citealt{Grether2006}). Building on this framework, we modified \texttt{TRICERATOPS} to compute the false positive probability of a candidate planet using transit light curves obtained in multiple bandpasses, and hereafter refer to this updated code as \texttt{TRICERATOPS+}\footnote{\url{https://github.com/JGB276/TRICERATOPS-plus}}. This code can support additional light curves in any of the SDSS $g$, $r$, $i$, or $z$ bands, as well as the 2MASS $J$, $H$, or $K$ bands. We accomplished this by adding an option in \texttt{TRICERATOPS} to calculate a separate log-likelihood for a light curve in a given bandpass. We then added this log-likelihood to the TESS log-likelihood (Equation 16 of \citealt{Giacalone2021}) when calculating the false positive probability. We note that the complete likelihood calculation includes not only the components described in Equation 16 of \citealt{Giacalone2021} but also an additional term, \texttt{lnprior\_companion}, which accounts for prior information about possible companions. This term is calculated using either user-provided contrast curves or the difference in TESS magnitude between the target star and companion. 

In order to calculate bandpass-specific log-likelihood, we must account for the effect of varying bandpasses on the shape of the model light curves. We did this by creating a table of the limb-darkening coefficients in each bandpass using the Python package \texttt{ExoTIC-LD} \citep{david_grant_2022_7437681}. Our new tables span the same grid of $T_\mathrm{eff}$ and log $g$ as the existing table of limb-darkening coefficients for the TESS bandpass. For scenarios involving eclipsing binaries and/or contamination from unresolved and resolved companions, the flux ratios between the target star and companions also depend on the bandpass. Thus, we also modified \texttt{TRICERATOPS} to calculate bandpass-specific flux ratios when computing model light curves. 

In the original implementation of \texttt{TRICERATOPS}, the flux ratio between the target star and a bound unresolved companion is determined using a relation that takes the mass of the companion star and converts it to a $\Delta$ TESS magnitude relative to a solar mass star. This relation was derived by drawing a spline relation through points in the TESS magnitude-stellar mass plane (Figure 4 of \citealt[][]{Giacalone2021}). These data points were acquired by querying the TIC for all stars located a distance between 99 and 101pc away. Subsequent versions of \texttt{TRICERATOPS} expanded this framework to calculate near-infrared flux ratios in the $J$, $H$, and $K$ filters. In \texttt{TRICERATOPS+}, we further expand this framework to calculate flux ratios in the optical SDSS $g$, $r$, $i$, and $z$ filters. Although we initially used the same methodology, we found that there was a scarcity of stars in the TIC with $M_* >$ 1 solar mass with $g$, $r$, $i$, and $z$ magnitudes. For this reason, we instead utilize a synthetic population of stars at 100 pc generated by TRILEGAL \citep{Girardi2005} to draw a spline in the magnitude-stellar mass plane. 

Once the flux ratio of the companion relative to a solar mass star is determined, TRICERATOPS calculates the flux ratio between the target and a solar mass star, and then uses these two ratios to calculate the flux ratio between the target star and companion star. If a TESS target does not have $g$, $r$, $i$, or $z$ magnitudes listed in the TIC and a user provides a light curve in any of these filters, \texttt{TRICERATOPS+} will estimate the target star's magnitude in these bands by applying photometric transformations from \citealt{Jester2005, Billir2005, Jordi2006, Bilir2008}. Alternatively, users can readily obtain these magnitudes by querying the PanSTARRS catalog \citep{Chambers2016, Flewelling2020} and then provide them as inputs to \texttt{TRICERATOPS+}.

In scenarios involving resolved Gaia stars that lie inside the TESS aperture \texttt{TRICERATOPS} uses 2D Gaussian PSFs to estimate the contamination to the TESS aperture \citep{Giacalone2021}. In \texttt{TRICERATOPS+}, if a resolved Gaia star also lies inside a ground-based aperture and is bright enough to host the transit signal, a user can specify for a given bandpass the fractional flux contribution of this star to their aperture. Alternatively, \texttt{TRICERATOPS+} will estimate the fractional flux contribution of a given bandpass using the available magnitudes on the TIC for the target star and the contaminating star.

Next, we used \texttt{TRICERATOPS+} to compute the False Positive Probability (FPP) and the Nearby False Positive Probability (NFPP) using the TESS data and contrast curve from high-resolution imaging. The FPP describes the probability that a transit signal does not originate from a planet transiting the target star. The NFPP describes the probability that the observed transit signal originates from a resolved nearby star (i.e., a star farther away than 2\arcsec~that contributes sufficient flux to the TESS aperture to produce the observed signal) rather than the target star. The FPP encapsulates all possible false positive scenarios, whereas the NFPP considers only a subset of them. Thus, the NFPP will always be less than or equal to the FPP. We then repeated the FPP and NFPP calculation after adding the Palomar $J$ band transit data and, if available, the LCOGT, FLWO/Keplercam, and MUSCAT2 data to quantify the change in FPP and NFPP. Following previous statistical validation studies with \texttt{TRICERATOPS} \citep[e.g.,][]{Giacalone2021, Giacalone2022}, we ran the analysis multiple times and report the average value and the 68$\%$ confidence interval of the FPP and the NFPP in Table \ref{tab:fpps}. Although \citealt{Giacalone2021} recommends running \texttt{TRICERATOPS} 20 times, we opted for 10 given our expanded dataset and increased computation time per run.

\section{Vetting and Disposition}
\label{sec:results}

In this section we present our analysis of each target, taking into account the full range of ground-based follow-up (high-resolution imaging, stellar spectroscopy, and multi-color transit photometry) available. We use radii (derived from the TESS light curve) and false positive probabilities to classify objects, with different classification criteria applied depending on whether the radius is smaller than 8 $R_\earth$ or larger than or equal to 8 $R_\earth$ (roughly equivalent to the minimum radius of a brown dwarf; \citealt{Sorahana2013}).  For candidates larger than 8 $R_\earth$, mass measurements are required in order to differentiate between bona fide planets and brown dwarfs. This means that objects larger than 8 $R_\earth$ cannot be validated as planets within the current statistical framework, since we cannot distinguish between planets and brown dwarfs using photometry alone \citep{Giacalone2021}.

For objects with radii smaller than 8 $R_\earth$, we use four classification categories based on False Positive Probability (FPP) and Nearby False Positive Probability (NFPP) as defined in \citealt{Giacalone2021}. The thresholds were chosen to reach a suitable level of confidence in a candidate's nature.
\begin{itemize}
    \item Validated Planet (VP): FPP $< 1.5\%$ and NFPP $< 0.1\%$
    \item Possible Planet (PP): $1.5\% \leq$ FPP $\leq 50\%$ and NFPP $< 0.1\%$
    \item False Positive (FP): FPP $> 70\%$ and/or NFPP $> 10\%$
    \item Inconclusive (I): All other cases.
\end{itemize}

For objects with radii greater than or equal to 8 $R_\earth$, we use only three categories:
\begin{itemize}
    \item Possible Planet (PP): FPP $\leq 50\%$ and NFPP $< 0.1\%$
    \item False Positive (FP): FPP $> 70\%$ and/or NFPP $> 10\%$
    \item Inconclusive (I): All other cases.
\end{itemize}

For candidates that are classified as false positives, we list the scenarios that are significantly contributing (relative probability > 1$\%$) to the overall false positive probability.

\begin{table*}[]
\centering
\fontsize{6.5pt}{6.5pt}\selectfont
\begin{tabular}{lllllllllll}
\hline
TOI & $t_0$-2450000 & $P$\tablenotemark{\scriptsize a} & $a/R_*$ & $b$ & $i$ & ${R_p/R_*}_\mathrm{WIRC}$ & ${R_p/R_*}_\mathrm{TESS}$ & $R_p$ & $T_\mathrm{eq}$  \\ 
 &  & (days) &  & & (deg) &  &  & ($R_\oplus$) & (K)
\\ \hline
1254.01 & $9601.99086_{-0.00025}^{+0.00025}$ & $1.018012_{-0.000056}^{+0.000056}$ & $2.571_{-0.093}^{+0.089}$ & $0.997_{-0.020}^{+0.036}$ & $67.18 \pm 1.09$ & $0.120_{-0.013}^{+0.026}$ & $0.086_{-0.013}^{+0.027}$ & - & - \\
\textbf{1346.01}\tablenotemark{\scriptsize b} & $9771.72743_{-0.00046}^{+0.00045}$ & $5.502558_{-0.000015}^{+0.000015}$ & $15.10_{-0.49}^{+0.23}$ & $0.16_{-0.11}^{+0.14}$ & $89.39 \pm 0.47$ & - & $0.02613_{-0.00035}^{+0.00036}$ & $2.23_{-0.14}^{+0.14}$ & 928 \\
\textbf{1346.02} & $9978.02390_{-0.00220}^{+0.00220}$ & $1.7622538_{-0.0000046}^{+0.0000047}$ & $7.42_{-0.42}^{+0.45}$ & $0.9261_{-0.0110}^{+0.0099}$ & $82.83 \pm 0.43$ & $0.0231_{-0.0063}^{+0.0048}$ & $0.02804_{-0.00075}^{+0.00077}$ & $2.39_{-0.16}^{+0.16}$ & 1324 \\
1616.01 & $9821.96350_{-0.00110}^{+0.00110}$ & $1.3432989_{-0.0000024}^{+0.0000024}$ & $3.92_{-0.16}^{+0.16}$ & $0.9605_{-0.0080}^{+0.0088}$ & $75.82 \pm 0.60$ & $0.0682_{-0.0074}^{+0.0079}$ & $0.0330_{-0.0016}^{+0.0024}$ & - & - \\
\textbf{2719.01} & $9882.95500_{-0.00480}^{+0.00530}$ & $3.37594_{-0.00005}^{+0.00005}$ & $5.68_{-0.32}^{+0.32}$ & $0.685_{-0.072}^{+0.055}$ & $83.07 \pm 0.76$ & $0.0308_{-0.0031}^{+0.0030}$ & $0.0332_{-0.0021}^{+0.0021}$ & $5.98_{-0.46}^{+0.46}$ & 1710 \\
4051.01 & $9639.01577_{-0.00079}^{+0.00078}$ & $1.537395_{-0.000003}^{+0.000003}$ & $4.44_{-0.28}^{+0.26}$ & $0.918_{-0.014}^{+0.019}$ & $78.07 \pm 0.77$ & $0.0961_{-0.0049}^{+0.0082}$ & $0.0559_{-0.0021}^{+0.0034}$ & - & - \\
4094.01 & $9897.63040_{-0.01330}^{+0.00320}$ & $4.911178_{-0.000055}^{+0.000056}$ & $10.45_{-0.48}^{+0.49}$ & $0.741_{-0.033}^{+0.030}$ & $85.93 \pm 0.26$ & $0.0301_{-0.0047}^{+0.0039}$ & $0.02420_{-0.00066}^{+0.00067}$ & - & - \\
\textbf{4155.01} & $9803.93840_{-0.00250}^{+0.00220}$ & $3.320729_{-0.000058}^{+0.000058}$ & $10.40_{-0.40}^{+0.27}$ & $0.21_{-0.14}^{+0.14}$ & $88.84 \pm 0.77$ & $0.0304_{-0.0053}^{+0.0045}$ & $0.03093_{-0.00071}^{+0.00070}$ & $3.21_{-0.17}^{+0.17}$ & 1255 \\
4731.01 & $9600.76920_{-0.00020}^{+0.00020}$ & $1.45536930_{-0.00000092}^{+0.00000089}$ & $4.68_{-0.14}^{+0.15}$ & $0.922_{-0.027}^{+0.040}$ & $78.64 \pm 0.55$ & $0.154_{-0.012}^{+0.022}$ & $0.148_{-0.012}^{+0.023}$ & - & - \\
5706.01 & $9822.64986_{-0.00035}^{+0.00036}$ & $0.3576493_{-0.0000074}^{+0.0000075}$ & $1.922_{-0.070}^{+0.075}$ & $1.065_{-0.047}^{+0.046}$ & $56.35 \pm 2.20$ & $0.136_{-0.038}^{+0.040}$ & $0.110_{-0.039}^{+0.042}$ & - & - \\
5735.01 & $9968.76949_{-0.00042}^{+0.00043}$ & $2.120839_{-0.000012}^{+0.000012}$ & $17.0_{-1.7}^{+2.2}$ & $1.003_{-0.050}^{+0.079}$ & $86.62 \pm 0.45$ & $0.087_{-0.028}^{+0.063}$ & $0.090_{-0.029}^{+0.064}$ & - & - \\
\textbf{6000.01} & $10109.9170_{-0.0014}^{+0.0012}$ & $0.448959_{-0.000011}^{+0.000011}$ & $4.66_{-0.14}^{+0.13}$ & $0.24_{-0.17}^{+0.18}$ & $87.05 \pm 2.16$ & $0.0253_{-0.0083}^{+0.0060}$ & $0.0228_{-0.0012}^{+0.0011}$ & $0.94_{-0.05}^{+0.05}$ & 1120 \\
\textbf{6324.01} & $10110.95431_{-0.00033}^{+0.00041}$ & $0.2792211_{-0.0000002}^{+0.0000002}$ & $4.29_{-0.69}^{+0.61}$ & $0.57_{-0.24}^{+0.16}$ & $82.36 \pm 2.94$ & $0.0306_{-0.0027}^{+0.0025}$ & $0.03217_{-0.00087}^{+0.00124}$ & $1.03_{-0.04}^{+0.04}$ & 1167 \\
6397.01 & $10098.68911_{-0.00050}^{+0.00051}$ & $1.670730_{-0.000001}^{+0.000001}$ & $6.39_{-0.25}^{+0.32}$ & $1.074_{-0.078}^{+0.081}$ & $80.32 \pm 0.84$ & $0.169_{-0.062}^{+0.071}$ & $0.179_{-0.061}^{+0.071}$ & - & - \\
\hline
\end{tabular}
\caption{Posterior results from the joint-fit of the TESS and Palomar light curves. The validated planets are marked in bold.}
\label{tab:posterior}
\vspace{0.1cm}
\footnotesize
\begin{flushleft}
\textbf{Notes.}
\tablenotemark{\scriptsize a} Derived from a separate transit fit to the TESS data.
\tablenotemark{\scriptsize b} Only TESS data was used in the analysis.
\end{flushleft}
\end{table*}

\subsection{TOI-1254.01 is a false positive}

Initial TESS pipeline vetting identified TOI-1254.01 (identified simultaneously by the QLP-FAINT and SPOC TPS pipelines; \citealt{Huang2020, Huang2020b, Kunimoto2021, Jenkins2002, Jenkins2010, Jenkins2020}) as a Jupiter-sized planet candidate ($R_p = 10.23$ $R_\earth$) orbiting the Sun-like star TOI-1254 ($V=$11.6; $J=$10.04, $T_\mathrm{eff}=5670$~K) every 1.018 days$^4$. Our observations show that TOI-1254.01 is a false positive. The high-resolution image from the NIRC2 instrument on Keck II in the $J$ filter shows no nearby stellar companions. The TESS data (sectors 14 - 75) in Figure \ref{fig:tess_obs} show a V-shaped light curve (${R_p/R_*}_\mathrm{TESS}$ + $b > $  1.04; \citealt{Thompson2018}), while the ground-based Palomar observations ($J$ band; Figure \ref{fig:palomar_obs}) and from LCOGT ($g$ and $z$ bands; Figure \ref{fig:lco_obs}) reveal that the transit is chromatic, with a deeper transit in $J$ and $z$ compared to the TESS band. The combination of a V-shaped light curve and wavelength-dependent transit depth strongly suggests that this candidate is a stellar eclipsing binary. This conclusion is further supported by our statistical analysis, which yields a high false positive probability (see Table \ref{tab:fpps}). Initially, using only the TESS light curve and the $J$ band contrast curve \texttt{TRICERATOPS+} yields a FPP of 0.9999 (68$\%$ confidence interval of 0.9999 to 0.9999) and a NFPP of 0.5066 (68$\%$ confidence interval of 0.1389 to 0.8743). When we incorporate the $J$ band Palomar light curve, the FPP remains at 0.9999. Similarly, when we incorporate the $g$ and $z$ LCOGT light curves, the FPP remains at 0.9999 (68$\%$ confidence interval of 0.9999 to 0.9999), but the NFPP decreases to zero because the stellar companion at 4.4\arcsec~(TIC 1102450081) that lies inside the optimal WIRC aperture is excluded from the $z$ LCOGT aperture. Of the 15 scenarios modeled by \texttt{TRICERATOPS+}, the SEBx2P scenario (an unresolved eclipsing binary with twice the orbital period around a secondary star) yields the highest relative probability (0.9999). In the Appendix, we plot the best-fit light curve for this scenario and compare it to the observations (see Figure \ref{fig:tricer_fits}).

\subsection{TOI-1346.01 is a validated planet}

TOI-1346.01 (identified by the QLP-FAINT pipeline; \citealt{Huang2020, Huang2020b, Kunimoto2021}) is a sub-Neptune planet candidate ($R_p = 2.23$ $R_\earth$) orbiting the relatively bright K star TOI-1346 ($V=$11.7; $J=$9.9, $T_\mathrm{eff}=5100$~K) every 5.503 days$^4$. Our observations indicate that TOI-1346.01 is a transiting planet. The high-resolution image from the NIRC2 instrument on Keck II in the $K$-continuum filter shows no nearby stellar companions. The TESS data (sectors 14 - 77) in Figure \ref{fig:tess_obs} show a U-shaped light curve (${R_p/R_*}_\mathrm{TESS}$ + $b < $  1.04; \citealt{Thompson2018}), and the ground-based light curves from LCOGT ($g$, $i$, $r$, and $z$ bands; Figure \ref{fig:lco_obs}) indicate that the light curve is achromatic. Our statistical analysis yields a low false positive probability (see Table \ref{tab:fpps}). We initially find a FPP of 0.0001 (68$\%$ confidence interval of 0.00003 to 0.0002) and a NFPP of 0.0001 (68$\%$ confidence interval of 0.00003 to 0.0001) for this candidate using the TESS data and the $K$-continuum contrast curve alone. When we incorporate the LCOGT light curves, the FPP decreases to $2 \times 10^{-14}$ (68$\%$ confidence interval of $2 \times 10^{-16}$ to $2 \times 10^{-12}$). Since there are no resolved nearby companions within the LCOGT apertures, the NFPP decreases to zero (i.e., we confirm that the transit must occur around the target star). With FPP$<$0.015 and NFPP$<10^{-3}$, TOI-1346.01 satisfies the criteria for a statistically validated planet and we hereafter refer to it as TOI-1346 c.

\subsection{TOI-1346.02 is a validated planet}

TOI-1346.02 (identified by the SPOC TPS pipeline; \citealt{Jenkins2002, Jenkins2010, Jenkins2020}) is a sub-Neptune planet candidate ($R_p = 2.39$ $R_\earth$) orbiting TOI-1346 every 1.762 days$^4$. It is interior to TOI-1346.01. Our observations indicate that TOI-1346.02 is a transiting planet, consistent with the observation that candidates in multi-planet systems are much more likely to be bona fide planets \citep{Rowe2014}. The high-resolution image from the NIRC2 instrument on Keck II in the $K$-continuum filter shows no nearby stellar companions. The TESS data (sectors 14 - 77) in Figure \ref{fig:tess_obs} show a U-shaped light curve (${R_p/R_*}_\mathrm{TESS}$ + $b < $  1.04; \citealt{Thompson2018}), and the ground-based light curves from Palomar ($J$ band; Figure \ref{fig:palomar_obs}) and LCOGT ($g$, $i$, and $z$ bands; Figure \ref{fig:lco_obs}) indicate that the light curve is achromatic, with only the $r$ LCOGT light curve showing a slightly deeper transit than TESS (greater than 1$\sigma$, but consistent within 2$\sigma$). Our statistical analysis yields a low false positive probability (see Table \ref{tab:fpps}). We initially find a FPP of $9.8 \times 10^{-4}$ (68$\%$ confidence interval of 0.00017 to 0.00179) and a NFPP of $7.2 \times 10^{-9}$ (68$\%$ confidence interval of $4.6 \times 10^{-9}$ to $9.8 \times 10^{-9}$) for this candidate using the TESS data and the $K$-continuum contrast curve alone. When we incorporate the Palomar light curve, the FPP decreases to $7.8 \times 10^{-6}$. Since there are no resolved nearby companions within the optimal WIRC aperture (3.5\arcsec), the NFPP decreases to zero (i.e., we confirm that the transit must occur around the target star). Lastly, when we incorporate the $g$, $r$, $i$, and $z$ LCOGT light curves, the FPP of this target decreases further from $7.8 \times 10^{-6}$ to $1.6 \times 10^{-7}$ (68$\%$ confidence interval of $2.6 \times 10^{-8}$ to $1 \times 10^{-6}$). With FPP$<$0.015 and NFPP$<10^{-3}$, TOI-1346.02 satisfies the criteria for a statistically validated planet and we hereafter refer to it as TOI-1346 b.

\subsection{TOI-1616.01 is a false positive}

Initial TESS pipeline vetting identified TOI-1616.01 (identified by the QLP-FAINT pipeline; \citealt{Huang2020, Huang2020b, Kunimoto2021}) as a Neptune-sized planet candidate ($R_p = 5.23$ $R_\earth$) orbiting the F star TOI-1616 ($V=$10.8; $J=$10.02, $T_\mathrm{eff}=6510$~K) with a period of 1.343 days$^4$. Our observations indicate that TOI-1616.01 is a false positive. The high-resolution image from SAI-2.5m in the $I$ filter shows no nearby stellar companions. The TESS data (sectors 24 - 77) in Figure \ref{fig:tess_obs} show a U-shaped light curve (${R_p/R_*}_\mathrm{TESS}$ + $b < $  1.04; \citealt{Thompson2018}), while the Palomar $J$ band observations in Figure \ref{fig:palomar_obs} indicate that the light curve is chromatic, with a deeper transit in the $J$ band compared to the TESS band. Our statistical analysis yields a high FPP (see Table \ref{tab:fpps}) of 0.7129 (68$\%$ confidence interval of 0.6628 to 0.7730 and a NFPP of 0.0122 (68$\%$ confidence interval of 0.0110 to 0.0134) based on the TESS photometry and the SAI-2.5m contrast curve alone. When we include the Palomar data, the FPP increases to 1 (68$\%$ confidence interval of 1 to 1). Since there are no resolved companions within the WIRC aperture (4.25\arcsec), the NFPP decreases to zero. Of the 15 scenarios modeled by \texttt{TRICERATOPS+}, the SEBx2P scenario (an unresolved eclipsing binary with twice the orbital period around a secondary star) yields the highest relative probability (1.0).

\subsection{TOI-2719.01 is a validated planet}

TOI-2719.01 (identified by the QLP-FAINT pipeline; \citealt{Huang2020, Huang2020b, Kunimoto2021}) is a Neptune-sized planet candidate ($R_p = 6.13$ $R_\earth$) orbiting the faint Sun-like star TOI-2719 ($V$=12.7; $J$=11.4, $T_\mathrm{eff}=5760$~K) every 3.38 days$^4$. The host star also appears to have a moderately enhanced metallicity of [m/H]$=0.27\pm0.08$. Our observations indicate that TOI-2719.01 is a transiting planet. The high-resolution image from the PHARO instrument on Palomar in the $K$-continuum filter reveals no nearby stellar companions. The TESS photometry (sector 32) in Figure \ref{fig:tess_obs} shows a flat-bottomed U-shaped transit (${R_p/R_*}_\mathrm{TESS}$ + $b < $  1.04; \citealt{Thompson2018}), and the Palomar observations (Figure \ref{fig:palomar_obs}) indicate an achromatic transit signal. Our statistical analysis yields a low false positive probability (see Table \ref{tab:fpps}). Initially, we find that the FPP is $1.9 \times 10^{-3}$ (68$\%$ confidence interval of 0.0015 to 0.0023) and that the NFPP is $4.1 \times 10^{-4}$ (68$\%$ confidence interval of 0.00040 to 0.00042) using only the TESS and the $K$ continuum contrast curve. When we incorporate the Palomar data, the FPP decreases to $7 \times 10^{-4}$ (68$\%$ confidence interval of $5 \times 10^{-4}$ to $9 \times 10^{-4}$). The NFPP decreases to zero, as there are no resolved nearby companions within the WIRC aperture (2.75\arcsec). With FPP$<$0.015 and NFPP$<10^{-3}$, TOI-2719 satisfies the criteria for a statistically validated planet and we hereafter refer to it as TOI-2719 b.

\subsection{TOI-4051.01 is a false positive}

Initial TESS pipeline vetting identified TOI-4051.01 (identified by the QLP-FAINT pipeline; \citealt{Huang2020, Huang2020b, Kunimoto2021}) as a Neptune-sized planet candidate ($R_p = 6.10$ $R_\earth$) orbiting an early K star ($V=$13.8; $J=$12.2, $T_\mathrm{eff}=5090$~K) every 1.537 days$^4$. Our observations indicate that TOI-4051.01 is a false positive. The high-resolution image from SAI-2.5m in the $I$ band shows a companion that is 0.53\arcsec~away from TOI-4051. The companion is 3.1 magnitudes fainter in the $I$ band. The TESS observations (sectors 47 - 74) in Figure \ref{fig:tess_obs} show a U-shaped transit signal (${R_p/R_*}_\mathrm{TESS}$ + $b < $  1.04; \citealt{Thompson2018}). The ground-based light curves from Palomar ($J$ band; Figure \ref{fig:palomar_obs}) and KeplerCam ($i$ band; Figure \ref{fig:lco_obs}) show a transit that is deeper in the $J$ and $i$ bands than in the TESS band. Our statistical analysis yields a high FPP of 0.9828 (68$\%$ confidence interval of 0.9734 to 0.9923) and a NFPP of $8 \times 10^{-14}$ (68$\%$ confidence interval of $1.7 \times 10^{-15}$ to $4.2 \times 10^{-12}$) using the TESS light curve and the SAI-2.5m contrast curve. When we incorporate the $J$ band Palomar light curve, the FPP increases to 1. The NFPP decreases to zero, as there are no resolved nearby stars within the WIRC aperture (2.5\arcsec). Lastly, when we incorporate the $i$ band light curve from KeplerCam, we find that the FPP remains at 1. Of the 15 scenarios modeled by \texttt{TRICERATOPS+}, the STP scenario (a transiting planet around a secondary unresolved star) and the SEBx2P scenario (an unresolved eclipsing binary with twice the orbital period around a secondary star) yield the highest relative probabilities (0.52 and 0.48, respectively).

\subsection{TOI-4094.01 may be a planet}

TOI-4094.01 (identified by the QLP-FAINT pipeline; \citealt{Huang2020, Huang2020b, Kunimoto2021}) is a sub-Neptune sized ($R_p = 3.27$ $R_\earth$) planet candidate orbiting an early G star ($V=$11.2; $J=$10.5, $T_\mathrm{eff}=6000$~K) every 4.911 days$^4$. Our observations indicate that TOI-4094.01 may be a transiting planet. The high-resolution image from SAI-2.5m in the $I$ band does not show evidence of a companion. The TESS observations (sectors 47 - 76) in Figure \ref{fig:tess_obs} show a U-shaped transit light curve (${R_p/R_*}_\mathrm{TESS}$ + $b < $  1.04; \citealt{Thompson2018}), but the partial $J$ band light curve in Figure \ref{fig:palomar_obs} reveals that TOI-4094 has a slightly chromatic transit signal where the transit is deeper in the $J$ band than in the TESS band. However, we find that the degree of chromaticity depends on the choice of detrending vectors in the systematics model. Our statistical analysis yields a somewhat low false positive probability (see Table \ref{tab:fpps}). We find a FPP of $4.6 \times 10^{-3}$ (68$\%$ confidence interval of 0.0017 to 0.014) and a NFPP of $7.1 \times 10^{-4}$ (68$\%$ confidence interval of 0.00027 to 0.0019) from the TESS data and the SAI-2.5m contrast curve alone. When we incorporate the Palomar $J$ band light curve, the FPP increases to 0.031 (68$\%$ confidence interval of 0.0157 to 0.1010) and the NFPP decreases to zero. However, the increase in FPP alone is enough to reclassify this candidate from a statistically validated planet to a possible planet. For this planet, a full infrared transit observation is needed to confirm or disprove the tentative chromaticity identified in our Palomar data.

\begin{table*}[ht]
\centering
\fontsize{5.5pt}{5.5pt}\selectfont
\begin{tabular}{lllllllll}
\hline
TOI & TCC & TCC & TCCM & TCCM & TCC & TCC & TCCM & D \\
 & FPP & FPP CI & FPP & FPP CI & NFPP & NFPP CI & NFPP &  \\
\hline
1254.01 & $99.99 \times 10^{-2}$ & ($97.96 \times 10^{-2}$, $1.000$) & $99.99 \times 10^{-2}$ & ($99.99 \times 10^{-2}$, $99.99 \times 10^{-2}$) & $50.66 \times 10^{-2}$ & ($13.89 \times 10^{-2}$, $87.43 \times 10^{-2}$) & $0$ & FP \\
1346.01 & $1.10 \times 10^{-4}$ & ($3.00 \times 10^{-5}$, $1.90 \times 10^{-4}$) & $2.00 \times 10^{-14}$ & ($2.00 \times 10^{-16}$, $2.00 \times 10^{-12}$) & $1.10 \times 10^{-4}$ & ($3.00 \times 10^{-5}$, $1.10 \times 10^{-4}$) & $0$ & VP \\
1346.02 & $9.80 \times 10^{-4}$ & ($1.70 \times 10^{-4}$, $1.79 \times 10^{-3}$) & $1.60 \times 10^{-7}$ & ($2.60 \times 10^{-8}$, $1.00 \times 10^{-6}$) & $7.20 \times 10^{-9}$ & ($4.60 \times 10^{-9}$, $9.80 \times 10^{-9}$) & $0$ & VP \\
1616.01 & $71.79 \times 10^{-2}$ & ($66.28 \times 10^{-2}$, $77.30 \times 10^{-2}$) & $1.000$ & ($1.000$, $1.000$) & $1.22 \times 10^{-2}$ & ($1.10 \times 10^{-2}$, $1.34 \times 10^{-2}$) & $0$ & FP \\
2719.01 & $1.90 \times 10^{-3}$ & ($1.50 \times 10^{-3}$, $2.30 \times 10^{-3}$) & $7.00 \times 10^{-4}$ & ($5.00 \times 10^{-4}$, $9.00 \times 10^{-4}$) & $4.10 \times 10^{-4}$ & ($4.00 \times 10^{-4}$, $4.20 \times 10^{-4}$) & $0$ & VP \\
4051.01 & $98.28 \times 10^{-2}$ & ($97.34 \times 10^{-2}$, $99.23 \times 10^{-2}$) & $1.000$ & ($1.000$, $1.000$) & $8.00 \times 10^{-14}$ & ($1.70 \times 10^{-15}$, $4.20 \times 10^{-12}$) & $0$ & FP \\
4094.01 & $4.57 \times 10^{-3}$ & ($1.70 \times 10^{-3}$, $1.40 \times 10^{-2}$) & $3.07 \times 10^{-2}$ & ($1.57  \times 10^{-2}$, $10.10  \times 10^{-2}$) & $7.06 \times 10^{-4}$ & ($2.66 \times 10^{-4}$, $1.89 \times 10^{-3}$) & $0$ & PP \\
4155.01 & $3.50 \times 10^{-3}$ & ($3.20 \times 10^{-3}$, $3.80 \times 10^{-3}$) & $1.20 \times 10^{-3}$ & ($1.00 \times 10^{-3}$, $1.40 \times 10^{-3}$) & $3.10 \times 10^{-4}$ & ($2.60 \times 10^{-4}$, $3.60 \times 10^{-4}$) & $0$ & VP \\
4731.01 & $61.63\times 10^{-2}$ & ($19.62\times 10^{-2}$, $90.20\times 10^{-2}$) & $3.01\times 10^{-2}$ & ($1.63 \times 10^{-3}$, $55.93 \times 10^{-2}$) & $1.00 \times 10^{-4}$ & ($2.00 \times 10^{-6}$, $6.00 \times 10^{-4}$) & 0 & I \\
5706.01 & $82.79 \times 10^{-2}$ & ($81.32 \times 10^{-2}$, $84.26 \times 10^{-2}$) & $99.99 \times 10^{-2}$ & ($99.99 \times 10^{-2}$, $99.99 \times 10^{-2}$) & $1.00 \times 10^{-25}$ & ($9.10 \times 10^{-26}$, $1.09 \times 10^{-25}$) & - & FP \\
5735.01 & $21.30 \times 10^{-2}$ & ($18.77 \times 10^{-2}$, $23.83 \times 10^{-2}$) & $20.20 \times 10^{-2}$ & ($18.40 \times 10^{-2}$, $22.00 \times 10^{-2}$) & $2.00 \times 10^{-10}$ & ($1.70 \times 10^{-10}$, $2.30 \times 10^{-10}$) & $0$ & PP \\
6000.01 & $7.06 \times 10^{-2}$ & ($6.98 \times 10^{-2}$, $7.14 \times 10^{-2}$) & $7.80 \times 10^{-3}$ & ($7.50 \times 10^{-3}$, $8.10 \times 10^{-3}$) & $3.84 \times 10^{-2}$ & ($3.77 \times 10^{-2}$, $3.91 \times 10^{-2}$) & $0$ & VP \\
6324.01 & $1.32 \times 10^{-8}$ & ($5.50 \times 10^{-10}$, $3.16 \times 10^{-7}$) & $8.51 \times 10^{-9}$ & ($3.80 \times 10^{-10}$, $1.91 \times 10^{-7}$) & $1.00 \times 10^{-16}$ & ($5.37 \times 10^{-17}$, $1.86 \times 10^{-16}$) & 0 & VP \\
6397.01 & $8.50 \times 10^{-2}$ & ($6.30 \times 10^{-2}$, $11.00 \times 10^{-2}$) & $1.50 \times 10^{-2}$ & ($4.00 \times 10^{-3}$, $2.56 \times 10^{-2}$) & $0$ & - & $0$ & PP \\
\hline
\end{tabular}
\caption{False positive probabilities for each candidate as calculated with \texttt{TRICERATOPS+}}
\label{tab:fpps}
\vspace{0.1cm}
\footnotesize
\begin{flushleft}
\textbf{Notes.}
TCC: TESS and contrast curve. TCCM: TESS, contrast curve, and multi-color. CI: Confidence interval. D: Disposition
\end{flushleft}
\end{table*}

\subsection{TOI-4155.01 is a validated planet}

TOI-4155.01 (identified by the QLP-FAINT pipeline; \citealt{Huang2020, Huang2020b, Kunimoto2021}) is a sub-Neptune sized ($R_p = 3.21$ $R_\earth$) planet candidate that orbits its Sun-like host star ($V=$11.9; $J=$10.6, $T_\mathrm{eff}=5720$~K) every 3.32 days$^4$. Our observations indicate that TOI-4155.01 is a transiting planet. The high-resolution image from WIYN-3.5m in the 832 nm filter shows no evidence of any nearby ($<1.2$\arcsec) stellar companions. The TESS photometry (sectors 52 - 59) in Figure \ref{fig:tess_obs} shows a flat-bottomed U-shaped transit light curve (${R_p/R_*}_\mathrm{TESS}$ + $b < $  1.04; \citealt{Thompson2018}), and the ground-based light curves from Palomar ($J$ band; Figure \ref{fig:palomar_obs}) and LCOGT (Figure \ref{fig:lco_obs}) confirm that the transit signal is achromatic. Our statistical analysis also yields a low false positive probability (see Table \ref{tab:fpps}). Using the TESS light curve and the WIYN-3.5m contrast curve alone, we find a FPP of $3.5 \times 10^{-3}$ (68$\%$ confidence interval of 0.0032 to 0.0038) and a NFPP of $3.1 \times 10^{-4}$ (68$\%$ confidence interval of $2.6 \times 10^{-4}$ to $3.6 \times 10^{-4}$). With the addition of the Palomar light curve in the $J$ band, we find that the FPP decreases to $1.2 \times 10^{-3}$ and that the NFPP decreases to zero. Lastly, with the addition of the $i$ LCOGT light curve, the FPP remains at $1.2 \times 10^{-3}$ (68$\%$ confidence interval of $1.0 \times 10^{-3}$ to $1.4 \times 10^{-3}$). With FPP$<$0.015 and NFPP$<10^{-3}$, TOI-4155.01 satisfies the criteria for a statistically validated planet and we hereafter refer to it as TOI-4155 b.

\subsection{TOI-4731.01 is inconclusive}

TOI-4731.01 (identified by the QLP-FAINT pipeline; \citealt{Huang2020, Huang2020b, Kunimoto2021}) is a Jupiter-sized ($R_p = 20$ $R_\earth$) planet candidate orbiting a late F star ($V$=12.4; $J$=11.3, $T_\mathrm{eff}=6010$~K) every 1.455 days$^4$. High-resolution imaging from SAI-2.5m in the $I$ filter reveals a companion (TIC 717538186) located 2.01\arcsec~away from TOI-4731. Although this companion appears 2.3 magnitudes fainter in the $I$ band, it is only 0.7 magnitudes fainter in the TESS band. This discrepancy likely arises from anisoplanatism, a known limitation of speckle interferometry that tends to overestimate brightness differences in widely separated binary systems.

Our ground-based observations suggest a few possible scenarios: TOI-4731.01 may be a giant transiting planet, a brown dwarf, or an eclipsing binary diluted by TIC 717538186. The TESS light curve (sectors 71 - 72) in Figure \ref{fig:tess_obs} shows a characteristic V-shaped transit (${R_p/R_*}_\mathrm{TESS}$ + $b > $  1.04; \citealt{Thompson2018}), and while the blended Palomar $J$ band light curve in Figure \ref{fig:palomar_obs} shows a shallower transit, the deblended light curve (using $J$=12.27 for TIC 717538186, estimated from its $V$ magnitude, TESS magnitude, and effective temperature) is consistent with the TESS observations at the 1$\sigma$ level (see $R_p/R_*$ in Table \ref{tab:posterior}). Similarly, the blended $g$, $r$, and $z$ MUSCAT2 light curves in Figure \ref{fig:lco_obs}, show that the transit signal is shallower in $g$, $r$ and $z$ than in the TESS band. The deblended MUSCAT2 light curves (see Section 2.5 for deblending details) yield radius ratios of $R_p/R_*$ ($g$) = $0.1284_{-0.0114}^{+0.0195}$, $R_p/R_*$ ($r$) = $0.1479_{-0.0126}^{+0.0198}$, and $R_p/R_*$ ($z$) = $0.1479_{-0.0101}^{+0.0156}$. 

\begin{figure*}
    \centering
    \includegraphics[width=0.95\textwidth]{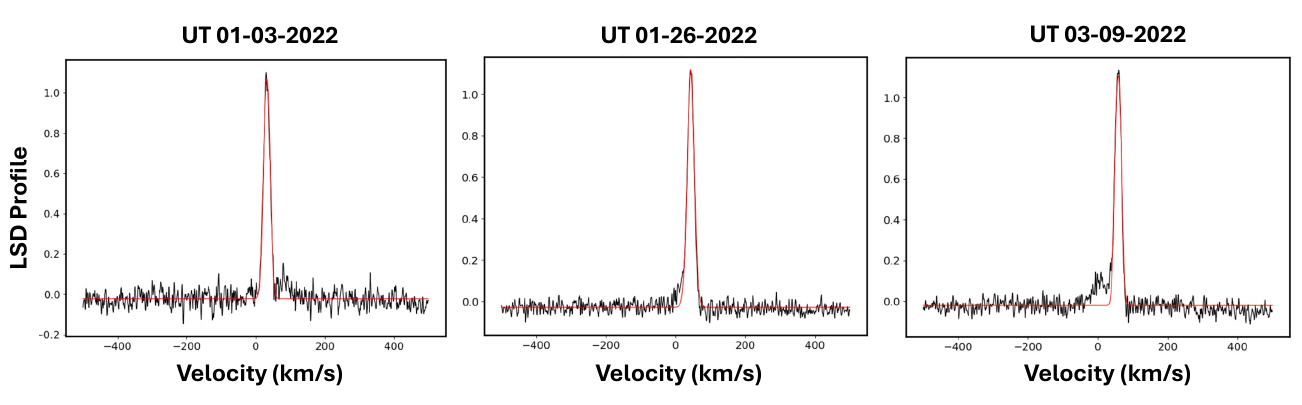}
    \caption{Least-Squares Deconvolution (LSD) profiles of the TRES stellar spectra of TOI-4731 showing the normalized line strength difference as a function of velocity. The velocity scale has been offset so that scattered moonlight occurs at \SI{0}{\kilo\meter\per\second}. The black line represents the observed profile, while the red line shows the best-fit model. The observations were taken on UT 2022-01-03, UT 2022-01-26, and UT 2022-03-09, corresponding to phases 31.7, 47.5, and 76.2, respectively, based on the TESS photometric ephemeris. Radial velocity measurements confirm that the primary peak remains stationary across all epochs (RV = 31.2, 30.2, and 30.7 \SI{}{\kilo\meter\per\second} at phases 31.7, 47.5, and 76.2, respectively), while a secondary peak exhibits clear orbital motion: red-shifted at +50 \SI{}{\kilo\meter\per\second} (phase 31.7), hidden behind the primary peak (phase 47.5), and blue-shifted at -50 \SI{}{\kilo\meter\per\second} (phase 76.2). This phase-dependent spectroscopic behavior, combined with V-shaped transits, indicates a potential hierarchical triple system where an inner eclipsing binary (responsible for the moving secondary peak) orbits a much brighter tertiary star (responsible for the stationary primary peak).}
    \label{fig:lsd}
\end{figure*}

Our TRICERATOPS analysis initially yielded an FPP of 0.61 and an NFPP of $1 \times 10^{-4}$ using only the TESS photometry and the SAI-2.5m contrast curve. Including the Palomar light curve reduces the FPP to 0.09, although with a wide 68$\%$ confidence interval of (0.0189, 0.6131), and reduces the NFPP to 0. When we incorporate the $g$, $r$, and $z$ MUSCAT2 light curves, we find that the FPP is further reduced to 0.03 and that the 68$\%$ confidence interval shrinks to ($1.6 \times 10^{-3}$, 0.5593). The FPP's broad uncertainty indicates that both of the modeled scenarios -- a transiting planet around TOI-4731 or an eclipsing binary -- can explain the observations. Our TRES spectrum for this target contains two sets of lines (Figure \ref{fig:lsd}), consistent with expectations for an unresolved eclipsing binary. However, the second component in the spectrum could also originate from contamination from the nearby star TIC 717538186. Given this ambiguity and the broad FPP confidence interval, we do not classify this system as a possible planet, despite its FPP$<50\%$ and NFPP$<0.1\%$.

\subsection{TOI-5706.01 is a false positive}

Initial TESS pipeline vetting identified TOI-5706.01 (identified simultaneously by the QLP-FAINT and SPOC TPS pipelines; \citealt{Huang2020, Huang2020b, Kunimoto2021, Jenkins2002, Jenkins2010, Jenkins2020}) as a sub-Neptune sized ($R_p = 2.61$ $R_\earth$) planet candidate that orbits its mid-K host star ($V=$11.6; $J=$9.8) every 0.358 days$^4$. Our observations indicate that TOI-5706.01 is a false positive. High-resolution imaging from Palomar in the $K$-continuum filter and Gemini at 832 nm reveal a companion 0.4\arcsec~away from TOI-5706. At 832 nm, the companion is 5.2 magnitudes fainter than TOI-5706, and in the $K$-continuum band it is 3.7 magnitudes fainter. Furthermore, the TESS observations (sectors 49 - 50) in Figure \ref{fig:tess_obs} show a V-shaped transit signal (${R_p/R_*}_\mathrm{TESS}$ + $b > $  1.04; \citealt{Thompson2018}). The ground-based light curves from Palomar ($J$ band; Figure \ref{fig:palomar_obs}) and LCOGT ($g$, $r$, $i$, and $z$; Figure \ref{fig:lco_obs}) further show that the transit is chromatic. Although the companion emits more flux at longer wavelengths and is located within the WIRC aperture (3\arcsec), the transit appears to be deeper in the $J$ band than in the TESS band. Our statistical analysis also yields a high false positive probability (see Table \ref{tab:fpps}). We initially calculate a FPP of 0.8279 (68$\%$ confidence interval of 0.8132 to 0.8426) and a NFPP of $1 \times 10^{-25}$ (68$\%$ confidence interval of $9.1 \times 10^{-26}$ to $1.1 \times 10^{-25}$) with the TESS data and the Gemini contrast curve. When we incorporate the $J$ band Palomar light curve the FPP increases to 0.9999, and when we incorporate the LCOGT light curves the FPP remains at 0.9999 (68$\%$ confidence interval of 0.9999 to 0.9999). However, we are unable to provide an updated NFPP since the flux ratio between the target and companion in the other bandpasses is unknown. Of the 15 scenarios modeled by \texttt{TRICERATOPS+}, the SEBx2P scenario (an unresolved eclipsing binary with twice the orbital period around a secondary star), the SEB scenario (an unresolved eclipsing binary with an orbital period of 0.3576 days around a secondary star), and the STP scenario (a transiting planet around a secondary unresolved star) yield the highest relative probabilities (0.83, 0.09, and 0.08, respectively).

\subsection{TOI-5735.01 may be a planet on a grazing orbit}

TOI-5735.01 (identified by the SPOC TPS pipeline; \citealt{Jenkins2002, Jenkins2010, Jenkins2020}) is a sub-Neptune sized ($R_p = 2.07$ $R_\earth$) planet candidate. It orbits its high proper motion M dwarf host star ($V=$15; $J=$10.6) every 2.121 days$^4$. The host star is the coolest in our sample with $T_\mathrm{eff}=3220$~K, resulting in an equilibrium temperature of 480 K$^4$ for the planet candidate. Our observations indicate that TOI-5735.01 may be a planet on a grazing orbit, although it remains possible that it is a false positive. High-resolution imaging from the NIRC2 instrument on Keck II in the $K$-continuum filter shows no evidence of any nearby ($<4$\arcsec) stellar companions. The TESS observations (sectors 14 - 74) in Figure \ref{fig:tess_obs} show a symmetric V-shaped transit light curve (${R_p/R_*}_\mathrm{TESS}$ + $b > $  1.04; \citealt{Thompson2018}), which frequently corresponds to false positives due to stellar eclipsing binaries. The ground-based light curves from LCOGT ($r$, $i$, and $z$; Figure \ref{fig:lco_obs}) and Palomar ($J$; Figure \ref{fig:palomar_obs}), reveal that the transit signal is achromatic in some bandpasses and chromatic in others. The transit depths are consistent across the TESS band, $r$, $z$, and $J$, but are shallower in $i$. We note that there are also two LCOGT light curves in $g$, but they are excluded from our analysis due to poor cadence. Our statistical analysis yields a somewhat high FPP of 0.2130 (68$\%$ confidence interval of 0.1877 to 0.2383) and a NFPP of $2 \times 10^{-10}$ (68$\%$ confidence interval of $1.7 \times 10^{-10}$ to $2.3 \times 10^{-10}$) using the TESS and contrast curve data (see Table \ref{tab:fpps}), probably as a result of the V-shaped transit light curve. With the addition of the Palomar light curve, we find a slightly lower FPP of 0.177. Since there are no resolved stars inside the WIRC aperture, the NFPP decreases to zero. Finally, with the addition of the LCOGT light curves, the FPP increases to 0.203 (68$\%$ confidence interval of 0.184 to 0.22). With FPP$<50\%$ and NFPP$<0.1\%$, TOI-5735.01 meets the criteria for a possible planet but is not statistically validated.

\subsection{TOI-6000.01 is a validated planet}

TOI-6000.01 (identified by the SPOC TPS pipeline; \citealt{Jenkins2002, Jenkins2010, Jenkins2020}) is an Earth-sized ($R_p = 0.95$ $R_\earth$) planet candidate. It orbits its faint M dwarf host ($V=$15.5; $J=$11.7, $T_\mathrm{eff}=3420$~K) TOI-6000 every 0.449 days$^4$. Our observations indicate that TOI-6000.01 is a transiting planet. The high-resolution image from SAI-2.5m in the $I$ filter shows no evidence of any nearby ($<1.7$\arcsec) stellar companions (see Figure \ref{fig:ccs}). Furthermore, the TESS observations (sectors 14 - 76) in Figure \ref{fig:tess_obs} show a U-shaped transit signal (${R_p/R_*}_\mathrm{TESS}$ + $b < $  1.04; \citealt{Thompson2018}). In Figure \ref{fig:palomar_obs}, the Palomar observations further show that the light curve is achromatic. Our statistical analysis also yields a low false positive probability (see Table \ref{tab:fpps}). Our initial FPP using the TESS data and contrast curve alone is 0.0706 (68$\%$ confidence interval of 0.0698 to  0.0714) and the corresponding NFPP is 0.0384 (68$\%$ confidence interval of 0.0377 to  0.0391), which places TOI-6000.01 in the possible planet regime. However, the addition of our Palomar light curve decreases the FPP to 0.0078 (68$\%$ confidence interval of 0.0075 to 0.0081) and the NFPP to 0. This means that the addition of the Palomar transit observation reclassified TOI-6000.01 from a possible planet to a statistically validated planet. Among our sample, this FPP reduction is one of the largest because the NFPP comprises a large fraction of the FPP. Since this planet candidate meets the criteria for validation, we refer to it as TOI-6000 b. 

\subsection{TOI-6324.01 is a validated planet}

TOI-6324.01 (identified by the QLP-FAINT pipeline; \citealt{Huang2020, Huang2020b, Kunimoto2021}) is an Earth sized ($R_p = 1.07R_\earth$) planet candidate that orbits an M dwarf ($V=$13.4; $J=$9.4, $T_\mathrm{eff}=3360$~K) every 0.279 days$^4$. Our observations indicate that TOI-6324.01 is a transiting planet. The planetary nature of this candidate was independently confirmed by radial velocity measurements, which revealed that it has a mass of 1.17 $\pm$ 0.22 $\mathrm{M_\earth}$ \citep{Lee2025}. The high-resolution image from the NIRC2 instrument on Keck II in the $K$-continuum filter shows no evidence of any nearby ($<4$\arcsec) stellar companions. However, according to the TIC there is a faint ($\Delta$Tmag $=$ 7) companion $0.9$\arcsec~away from TOI-6324 that is also located inside the optimal WIRC aperture (3.75\arcsec). The TESS observations (sectors 16 - 58) show a flat-bottomed U-shaped transit light curve (${R_p/R_*}_\mathrm{TESS}$ + $b < $  1.04; \citealt{Thompson2018}), which the Palomar observations confirm is achromatic despite the contamination from the faint companion. Our statistical analysis also yields a low false positive probability (see Table \ref{tab:fpps}). We initially find a FPP of $1.3 \times 10^{-8}$ (68$\%$ confidence interval of $5.5 \times 10^{-10}$ to $3.2 \times 10^{-7}$) and a NFPP of $1 \times 10^{-16}$ (68$\%$ confidence interval of $5.4 \times 10^{-17}$ to $1.9 \times 10^{-16}$) using the TESS data and the contrast curve alone. When we include the Palomar light curve, we find that the FPP decreases to $8.5 \times 10^{-9}$ (68$\%$ confidence interval of $3.8 \times 10^{-10}$ to $1.9 \times 10^{-7}$) and that the NFPP decreases to zero. Since this planet candidate meets the criteria for validation, we hereafter refer to it as TOI-6324 b.

\subsection{TOI-6397.01 may be a planet}

TOI-6397.01 (identified by the QLP-FAINT pipeline; \citealt{Huang2020, Huang2020b, Kunimoto2021}) is a giant ($R_p = 18.07$ $R_\earth$) planet candidate that orbits an early K star ($V=$13.9; $J=$12.4, $T_\mathrm{eff}=5040$~K) every 1.671 days$^4$. Our observations indicate that TOI-6397.01 is a possible transiting planet. High-resolution imaging from SAI-2.5m in the $I$ filter shows no evidence of any nearby ($<1$\arcsec) stellar companions. The TESS light curve (sectors 73 - 81) is V-shaped (${R_p/R_*}_\mathrm{TESS}$ + $b > $  1.04; \citealt{Thompson2018}), and the Palomar light curve in $J$ indicates that the transit is slightly chromatic. We find that the transit is slightly deeper in the TESS band than in the $J$ band. Our initial statistical analysis returns a moderate FPP (see Table \ref{tab:fpps}) of 0.085 (68$\%$ confidence interval of 0.063 to 0.11) and a NFPP of 0 using the TESS light curve and the contrast curve alone. When we include the Palomar light curve, the FPP decreases to 0.015 (68$\%$ confidence interval of 0.004 to 0.026) and confirms that the NFPP is zero, as there are no companions inside the optimal WIRC aperture (3.5\arcsec). With a radius > 8 $R_\earth$, FPP $<50\%$ and NFPP $<0.1\%$, TOI-6371.01 is classified as a possible planet.

\section{Stellar and Planetary Characterization}
\label{sec:results2}

\begin{figure*}
    \centering
    \includegraphics[width=0.95\textwidth]{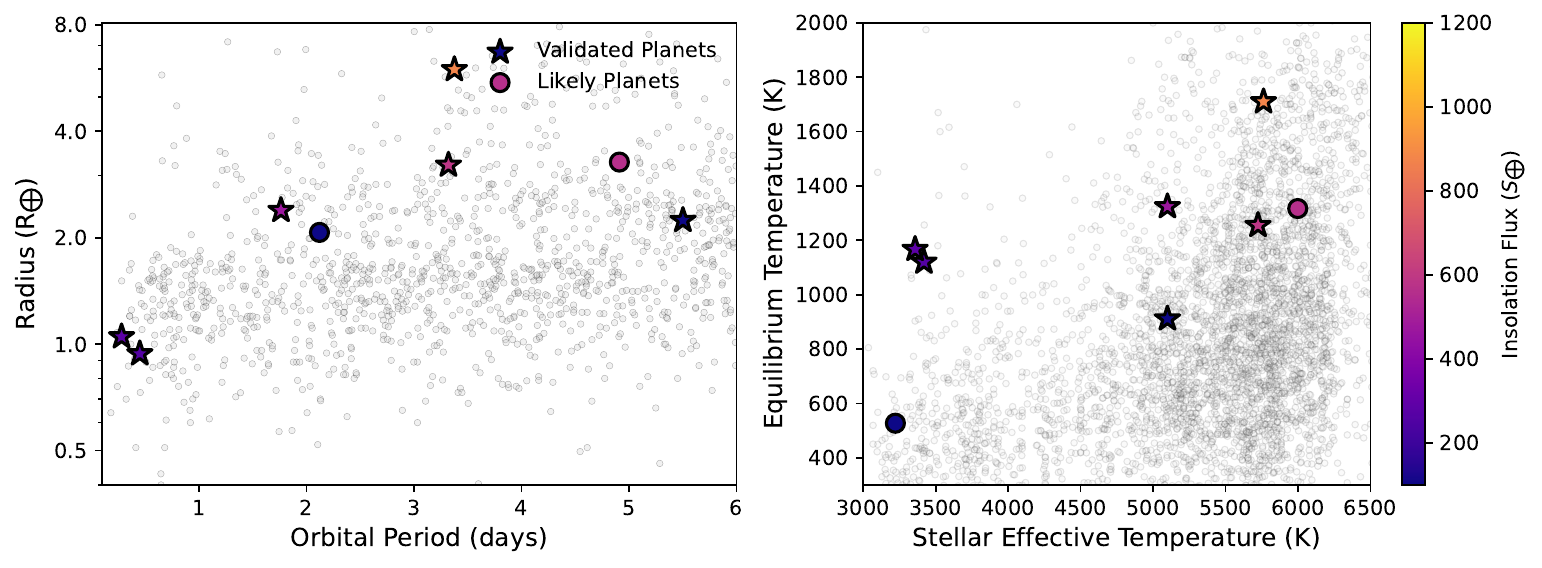}
    \caption{The properties of the validated systems (stars) and possible planets with radii smaller than 8 $R_\earth$ (circles) from our sample compared to the confirmed systems (gray circles) from the NASA Exoplanet Archive. The validated systems and possible planets are color coded based on their insolation flux.}
    \label{fig:sample}
\end{figure*}

Having validated the planetary nature of 6/14 candidates, we now present the stellar and planetary properties of these newly confirmed exoplanets. Table \ref{tab:stellar_props} summarizes the stellar parameters, including effective temperature and metallicity (from TRES spectra when available), and stellar radius and mass (from TIC). Table \ref{tab:posterior} details the planetary parameters, including the planet radius (derived from $R_*$ and the combined WIRC and TESS planet-to-star radius ratios) and the equilibrium temperature (calculated assuming zero albedo and full day-night heat redistribution). Figure \ref{fig:sample} illustrates the properties of the systems compared to the confirmed planets in the NASA Exoplanet Archive \citep{Akeson2013, ps}. 

Our validated systems exhibit diverse stellar and planetary characteristics:
\begin{itemize}
    \item TOI-2719 b and TOI-4155 b are sub-Neptunes orbiting Sun-like stars with orbital periods of approximately 3.3 days.
    \item TOI-1346 b and TOI-1346 c are sub-Neptunes orbiting a star slightly cooler than the Sun.
    \item TOI-6000 b and TOI-6324 b are Earth-sized planets on ultra-short-period orbits around M-type stars.
\end{itemize}

\section{Discussion}
\label{sec:discussion}

\subsection{What kinds of planet candidates benefit from multi-color photometry?}

We quantify the impact of our multi-color light curves on the false positive probabilities of TESS planet candidates in Table \ref{tab:fpps}. In general, we find that multi-color transit observations are most valuable for candidates in the possible planet category (FPP between $1.5-50\%$). Our results show that the addition of multi-color transit observations can reduce the FPPs of candidates with values near the statistical validation threshold of 1.5$\%$, reclassifying them as statistically validated planets. For example, our initial FPP for TOI-6000 b using the TESS light curve and the contrast curve alone was 0.071, which is insufficient for statistical validation. After incorporating our Palomar $J$ band light curve, we reduced the FPP to 0.0078. Conversely, the tentative chromaticity observed in our partial transit observation of TOI-4094.01 is enough to increase the FPP for this candidate from 0.005 to 0.031, reclassifying it from a statistically validated planet to a possible planet candidate.

While multi-color transit photometry is useful for confirming or disproving candidates in the possible planet category of FPP, it also serves another important purpose. The NFPP quantifies the likelihood of false positive scenarios where the transit signal originates from resolved stars other than the target star that lie within the TESS aperture. Ground-based observations typically offer higher angular resolution than TESS, enabling observers to resolve and exclude light from nearby stars that are blended within the TESS aperture. For candidates with non-zero NFPPs, which are more likely to orbit faint host stars, ground-based photometry can confirm that the transit occurs around the primary star and reduce the corresponding NFPP to zero (e.g, TOI-1254.01, TOI-1346 b, TOI-1346 c). If there is a companion star located inside both the TESS and ground-based apertures (e.g., TOI-6324 b), we can measure the chromaticity of the transit signal and confirm that it is consistent with a transiting planet orbiting the primary star diluted by light from the background star. However, in order to do so we must have some basic knowledge of the background star's radius and effective temperature in order to scale the measured flux ratio to different bandpasses. Without this information, we cannot calculate an updated NFPP incorporating the information from our multi-color photometry data.

\subsection{What kind of follow-up is most helpful for reducing the FPP and NFPP?}

Our updated version of \texttt{TRICERATOPS} incorporates two types of ground-based observations: contrast curves and multi-color light curves in the SDSS $g$, $r$, $i$, $z$ or 2MASS $J$, $H$, $K$ filters. Contrast curves provide significant constraints on the properties of simulated stars in various false positive scenarios, often leading to substantial reductions in the FPP. To illustrate this, we recalculated the FPP of TOI-4155 b with and without a contrast curve. Including the TESS data and the contrast curve yielded a FPP of 0.0035, while excluding the contrast curve increased the FPP to 0.0106. Notably, contrast curves typically require only a fraction of the observation time needed to acquire a full transit light curve. As a result, we recommend prioritizing high-resolution imaging for all candidates with moderate to low FPPs as an initial step in the validation process followed by multi-color transit photometry for candidates that remain in the possible planet category.

Although ground-based light curves in separate filters are also valuable, their impact on the FPP can vary depending on several factors. The effectiveness of a light curve in reducing the FPP will depend on the chosen filter, the spectral type of the host star, and whether the observed transit is chromatic or achromatic relative to the TESS observations.

Chromatic light curves (i.e., those showing deeper or shallower transits than TESS) can significantly increase the FPP of a planet candidate, regardless of the filter used. For example, in the case of TOI-5706.01, the LCOGT and Palomar observations revealed that the transit depth is wavelength-dependent. When we incorporated the $i$ or $z$ band light curve into our \texttt{TRICERATOPS} analysis along with the TESS data and contrast curve, the FPP increased to 0.9998 and 0.9999, respectively. This increase is comparable to that observed with the incorporation of a $J$ band light curve from Palomar.

For achromatic light curves, the impact on the FPP varies depending on the filter used and the host star's spectral type. We explored the importance of filter choice by simulating achromatic light curves for a planet candidate orbiting either a Sun-like star or an M dwarf in all four SDSS filters ($g$, $r$, $i$, $z$) and the $J$, $H$, and $K$ filters available in \texttt{TRICERATOPS+}. To simulate these light curves, we used \texttt{TRICERATOPS}'s transiting planet light curve model, adopting the appropriate limb darkening coefficients for each bandpass while holding all other transit shape parameters constant across filters. We kept the noise fixed to the same value across all bands in order to isolate the effect of bandpass on our results. For the Sun-like case, the light curves had an error bar of 479 ppm and for the M-type star case, they had an error bar of 279 ppm. These uncertainties were derived from the precision of the TESS light curves of TOI-2719 b (Sun-like star) and TOI-6000 b (M-type star) at the same cadence as the simulated light curves. We note that we did not perturb the simulated points away from the nominal values, but instead just assigned error bars to these nominal points. Among the available filters, $g$ band provides the greatest contrast with the TESS filter, and we find that it is generally the most effective in ruling out chromatic false positive scenarios and reducing the FPP. After the $g$ band, the effectiveness decreases from $r$ to similarly performing bands $J$, $H$, and $K$, and finally to $i$ and $z$. The effect of the $g$ band is particularly pronounced for M-type stars, where a $g$ band light curve can have a substantially greater impact compared to other filters (assuming the same signal-to-noise ratio). For G-type stars, a $g$ band light curve is still the most effective but the impact of other filters is more comparable. In reality, the achievable signal-to-noise ratio for a ground-based transit observation can vary significantly depending on the choice of bandpass. Although $g$ band observations are potentially the most impactful for M dwarf planet candidates, the relative faintness of these stars at shorter wavelengths means that the signal-to-noise ratio is often lower than at longer wavelengths.

\section{Conclusion}
\label{sec:conclusion}

In this paper, we present an updated version of \texttt{TRICERATOPS}, a popular tool for statistically validating planet candidates. Our updated \texttt{TRICERATOPS+}$^6$ package now supports the integration of multiple ground-based light curves when computing a planet candidate's false positive probability. We test this improved framework by applying it to 14 TESS planet candidates, utilizing primarily $J$ band light curves obtained with the 200-inch Hale Telescope at Palomar Observatory together with complementary archival observations from the Las Cumbres Observatory Global Telescope, Fred Lawrence Whipple Observatory, and the Teide Observatory. We statistically validate six new planets (TOI-1346 b, TOI-1346 c, TOI-2719 b, TOI-4155 b, TOI-6000 b, and TOI-6324 b) and explore the relative importance of multi-wavelength transit photometry and high-resolution imaging in the candidate validation process. 

Our analysis demonstrates the varying impact of different types of follow-up observations on the calculated false positive probability. Although contrast curves from high-resolution imaging consistently provide valuable constraints, the impact of multi-color transit light curves on the FPP is more complex and depends on factors such as the observed chromaticity relative to TESS observations, the specific filter used, and the host star's spectral type. Given these insights, \texttt{TRICERATOPS+} now serves as both a robust validation tool and a powerful planning tool. Using simulated light curves, our updated framework can be used to quantify the potential impact of various proposed multi-color observations, allowing for better optimization of observing strategies.

\section{Acknowledgments}

We thank the anonymous referee for a helpful report. This work is based on observations obtained at the Hale Telescope, Palomar Observatory, as part of a collaborative agreement between the Caltech Optical Observatories and the Jet Propulsion Laboratory (operated by Caltech for NASA). We are grateful to the Palomar Observatory staff and telescope operators, including Kathleen Koviak, Paul Nied, Tom Barlow, Carolyn Heffner, Isaac Wilson, Diana Roderick, and Joel Pearman, for their invaluable assistance with the observations. This research was carried out at the Jet Propulsion Laboratory and the California Institute of Technology under a contract with the National Aeronautics and Space Administration (80NM0018D0004) and funded through the President's and Director's Research $\&$ Development Fund Program.

This paper made use of data collected by the TESS mission. These data are publicly available from the Mikulski Archive for Space Telescopes (MAST) operated by the Space Telescope Science Institute (STScI). Funding for the TESS mission is provided by NASA's Science Mission Directorate. We acknowledge the use of public TESS data from pipelines at the TESS Science Office and at the TESS Science Processing Operations Center. Resources supporting this work were provided by the NASA High-End Computing (HEC) Program through the NASA Advanced Supercomputing (NAS) Division at Ames Research Center for the production of the SPOC data products. KAC and CNW acknowledge support from the TESS mission via subaward s3449 from MIT.

This work makes use of observations from the LCOGT network. Part of the LCOGT telescope time was granted by NOIRLab through the Mid-Scale Innovations Program (MSIP). MSIP is funded by NSF. This paper is based on observations made with the MuSCAT instruments, developed by the Astrobiology Center (ABC) in Japan, the University of Tokyo, and Las Cumbres Observatory (LCOGT). MuSCAT3 was developed with financial support by JSPS KAKENHI (JP18H05439) and JST PRESTO (JPMJPR1775), and is located at the Faulkes Telescope North on Maui, HI (USA), operated by LCOGT. MuSCAT4 was developed with financial support provided by the Heising-Simons Foundation (grant 2022-3611), JST grant number JPMJCR1761, and the ABC in Japan, and is located at the Faulkes Telescope South at Siding Spring Observatory (Australia), operated by LCOGT. This article is based on observations made with the MuSCAT2 instrument, developed by ABC, at Telescopio Carlos S'{a}nchez operated on the island of Tenerife by the IAC in the Spanish Observatorio del Teide.

The work of HPO has been carried out within the framework of the NCCR PlanetS supported by the Swiss National Science Foundation under grants 51NF40-182901 and 51NF40-205606. This work is partly financed by the Spanish Ministry of Economics and Competitiveness through grants PGC2018-098153-B-C31. This work is partly supported by JSPS KAKENHI Grant Numbers JP24H00017, JP24K00689, and JSPS Bilateral Program Number JPJSBP120249910.

We acknowledge financial support from the Agencia Estatal de Investigaci\'on of the Ministerio de Ciencia e Innovaci\'on MCIN/AEI/10.13039/501100011033 and the ERDF “A way of making Europe” through project PID2021-125627OB-C32, and from the Centre of Excellence “Severo Ochoa” award to the Instituto de Astrofisica de Canarias. F. M. acknowledges the financial support from the Agencia Estatal de Investigaci\'{o}n del Ministerio de Ciencia, Innovaci\'{o}n y Universidades (MCIU/AEI) through grant PID2023-152906NA-I00.

This research has made use of the Exoplanet Follow-up Observation Program (ExoFOP; DOI: 10.26134/ExoFOP5) website, which is operated by the California Institute of Technology, under contract with the National Aeronautics and Space Administration under the Exoplanet Exploration Program.

\newpage
\appendix

In Tables \ref{tab:palomar_obs}, \ref{tab:lco_keplercam_obs}, and \ref{tab:imaging}, we provide details of the ground-based light curve and high-resolution imaging observation of our targets.

\begin{table*}[h]
\centering
\begin{tabular}{cccccccccccc}
\hline
TOI & Date & Start Time & End Time & $t_{\exp}$ & $n_\mathrm{star}$\tablenotemark{\scriptsize a} & $r_\mathrm{phot}$\tablenotemark{\scriptsize b} & $d_\mathrm{comp}$\tablenotemark{\scriptsize c} & $\Delta \, \mathrm{T}$\tablenotemark{\scriptsize d} & Airmass & $\sigma/$10-min\tablenotemark{\scriptsize e} & $\sigma/\sigma_\mathrm{phot}$\tablenotemark{\scriptsize f} \\
& (UTC)& & & (s) & & (\arcsec) & (\arcsec) & & Start/Middle/End & (ppm) & \\
\hline
1254.01 & 2022-01-22 & 10:29:16 & 13:56:12 & 20 & 3 & 5.25 & 4.40 & 5.4 & 1.73/1.35/1.23 & 468 & 7.7 \\
1346.02 & 2023-02-02 & 11:39:58 & 13:38:38 & 24 & 4 & 3.50 & 33.60 & 6.5 & 1.57/1.42/1.32 & 133 & 2.1 \\
1616.01 & 2022-08-30 & 10:34:36 & 12:29:05 & 26.4 & 6 & 4.25 & 7.70 & 5.4 & 1.47/1.62/1.84 & 297 & 4.5 \\
2719.01 & 2022-10-30 & 07:41:13 & 13:32:35 & 48 & 10 & 2.75 & 11.60 & 5.5 & 1.50/1.22/1.95 & 367 & 3.0 \\
4051.01 & 2022-02-28 & 11:23:08 & 13:21:47 & 30 & 5 & 2.50 & 36.60 & 4.8 & 1.32/1.28/1.26 & 273 & 1.7 \\
4094.01 & 2022-11-14 & 02:18:14 & 06:30:22 & 30 & 11 & 3.00 & 23.40 & 6.9 & 1.30/1.55/2.08 & 300 & 4.7 \\
4155.01 & 2022-08-12 & 10:00:02 & 12:19:26 & 36 & 10 & 3.00 & 12.70 & 9.1 & 1.44/1.52/1.62 & 202 & 2.4 \\
4731.01 & 2022-01-21 & 05:32:29 & 08:21:33 & 20 & 5 & 3.75 & 2.00 & 0.7 & 1.08/1.06/1.19 & 470 & 4.3 \\
5706.01 & 2022-08-31 & 03:00:51 & 04:42:22 & 16 & 4 & 3.00 & 17.70 & 7.3 & 1.25/1.37/1.56 & 291 & 4.3 \\
5735.01 & 2023-01-24 & 05:26:35 & 10:10:16 & 25 & 3 & 3.75 & 20.60 & 5.2 & 1.60/1.42/1.40 & 271 & 3.5 \\
6000.01 & 2023-06-14 & 08:49:46 & 12:01:17 & 30 & 12 & 2.25 & 13.30 & 6.0 & 1.23/1.22/1.30 & 172 & 1.3 \\
6324.01 & 2023-06-15 & 09:30:57 & 11:50:55 & 10.5 & 6 & 3.75 & 0.90 & 7.0 & 1.34/1.25/1.21 & 157 & 2.4 \\
6397.01 & 2023-06-03 & 03:52:50 & 05:32:03 & 45 & 2 & 3.50 & 31.40 & 6.7 & 2.04/1.72/1.50 & 874 & 4.8 \\
\hline
\end{tabular}
\caption{Palomar observation details}
\label{tab:palomar_obs}
\vspace{0.05cm}
\footnotesize
\begin{flushleft}
\textbf{Notes.} 
\tablenotemark{\scriptsize a} Number of comparison stars used in detrending. 
\tablenotemark{\scriptsize b} Radius of the optimal photometric aperture.
\tablenotemark{\scriptsize c} Separation of the nearest star in the TIC.
\tablenotemark{\scriptsize d} The difference in TESS magnitudes between the target star and the nearest star in the TIC.
\tablenotemark{\scriptsize e} The 10-minute bin precision.
\tablenotemark{\scriptsize f} The 10-minute bin precision divided by the photon noise.
\end{flushleft}
\end{table*}

\begin{table*}[h]
\centering
\begin{tabular}{ccccccccc}
\hline
TOI & Observatory & Ap & Date  & Start Time & End Time & Filter(s) & $t_{\exp}^g$ & Airmass \\
 &  & (m) & (UTC) &  &  &  & (s) & Start/Middle/End \\
\hline

1254.01 & LCO-McD$^a$ & 1.0 & 2021-05-26 & 03:31:58 & 06:57:44 & $g', z_s$$^b$ & 19/45 & 1.41/1.25/1.22 \\
1346.01 & LCO-TEID$^c$ & 1.0 & 2024-05-17 & 22:21:17 & 03:34:34 & $i'$ & 30 & 1.69/1.36/1.34\\
1346.01 & LCO-HAL-M3$^d$ & 2.0 & 2023-05-26 & 6:44:43 & 11:31:25 & $g',r', i', z_s$ & 62/20/24/38 & 2.19/1.63/1.50\\
%1346.02 & LCO-TEID  & 1.0 & 2021-09-05 & 21:11:00 & 23:44:51 & $i'$ & 30 & 1.39/1.51/1.78 \\
%1346.02 & LCO-TEID  & 1.0 & 2022-05-22 & 22:27:12 & 01:23:30 & $i'$ & 30 & 1.61/1.35/1.33 \\
%1346.02 & LCO-TEID  & 1.0 & 2022-04-07  & 02:36:57 & 05:32:26 & $i'$ & 30 & 1.45/1.34/1.32 \\
1346.02 & LCO-HAL-M3  & 2.0 & 2022-03-20 & 11:41:07 & 14:35:20 & $g', r', i', z_s$ & 62/20/24/38 & 1.99/1.68/1.53 \\
%1346.02 & LCO-TEID  & 1.0 & 2021-09-05 & 21:11:00 & 23:44:51 & $i'$ & 30 & 1.39/1.51/1.78 \\
1346.02 & LCO-HAL-M3  & 2.0 & 2021-07-04 & 10:52:38 & 13:13:00 & $g', r', i', z_s$ & 62/20/24/38 & 1.63/1.82/2.19 \\
%1346.02 & LCO-McD  & 1.0 & 2021-04-23 & 05:22:00 & 07:37:23 & $i'$ & 30 & 1.70/1.47/1.37 \\
%1346.02 & LCO-McD  & 1.0 & 2021-04-07 & 08:08:00 & 10:58:35 & $i'$ & 30 & 1.42/1.31/1.27 \\
4051.01 & FLWO-KeplerCam$^e$ & 1.2 & 2024-05-20 & 03:20:23 & 08:30:55 & $i'$ & 38 & 1.61/1.34/1.31 \\
4155.01 & LCO-TEID  & 1.0 & 2022-08-18  & 23:08:00 & 04:10:55 & $i'$ & 17 & 1.59/1.55/1.70 \\
4731.01 & TEID-MUSCAT2$^f$	&  1.5 &	2022-02-14 &  20:02:42 & 01:05:56 &  $g', r', z_s$  & 20/7/25 &  1.16/1.04/1.39 \\
5706.01 & LCO-HAL-M3 & 2.0 & 2023-03-19 & 09:20:31 & 11:38:51 & $g', r', i', z_s$ & 68/20/24/38 & 2.00/1.59/1.37\\
5735.01 & LCO-HAL-M3 & 2.0 & 2024-02-18 & 10:29:47 & 13:31:17 & $g', r', i', z_s$ & 337/51/25/22 & 1.81/1.86/2.05\\
5735.01 & LCO-HAL-M3 & 2.0 & 2024-02-01 & 11:08:35 & 14:15:46 & $g', r', i', z_s$ & 337/51/25/22 & 1.81/1.84/2.00\\

\hline
\end{tabular}
\caption{LCOGT, KeplerCam, and MUSCAT 2 observation details}
\label{tab:lco_keplercam_obs}
\vspace{0.05cm}
\footnotesize
\begin{flushleft}
\textbf{Notes.} \tablenotemark{\scriptsize a} McDonald Observatory near Fort Davis, Texas, United States (McD). 
\tablenotemark{\scriptsize b} Pan-STARRS $z_s$ band. 
\tablenotemark{\scriptsize c} Teide Observatory on the island of Tenerife (TEID)
\tablenotemark{\scriptsize d} MuSCAT3 multi-band imager\citep{Narita:2020} on the 2\,m Faulkes Telescope North at Haleakala Observatory on Maui, Hawai'i.
\tablenotemark{\scriptsize e} Fred L. Whipple Observatory / KeplerCam.
\tablenotemark{\scriptsize f} Telescopio Carlos S\'{a}nchez (TCS) at Teide Observatory/MUSCAT2.
\tablenotemark{\scriptsize g} Exposure time in seconds. Multiple values indicate different exposure times used for simultaneous observations in separate filters, with longer exposure times typically employed for filters where the star appears dimmer to maintain adequate signal-to-noise ratio across all wavelengths.
\end{flushleft}
\end{table*}

\begin{table*}[h]
\centering
\fontsize{8pt}{8pt}\selectfont
\begin{tabular}{lllll}
\hline
TOI & Telescope & Instrument & Filter & Image Type \\ \hline
1254 & & & & \\ \hline
& \textbf{Keck2 (10 m)} & \textbf{NIRC2} & \textbf{J} & \textbf{AO} \\ \hline
& SAI-2.5m (2.5 m) & Speckle Polarimeter & I & Speckle \\ \hline
& Keck2 (10 m) & NIRC2 & K & AO \\ \hline
& WIYN (3.5 m) & NESSI & 832 (40) nm & Speckle \\ \hline
& WIYN (3.5 m) & NESSI & 562 (44) nm & Speckle \\ \hline
1346 & & & & \\ \hline
& WIYN (3.5 m) & NESSI & 562 (44) nm & Speckle \\ \hline
& WIYN (3.5 m) & NESSI & 832 (40) nm & Speckle \\ \hline
& \textbf{Keck2 (10 m)} & \textbf{NIRC2} & \textbf{K} & \textbf{AO} \\ \hline
& Gemini (8 m) & NIRI & Brgamma & AO \\ \hline
& 2.2m@CAHA (2.2 m) & AstraLux & SDSSz & Lucky \\ \hline
& Gemini (8 m) & NIRI & K & AO \\ \hline
& SAI-2.5m (2.5 m) & Speckle Polarimeter & I & Speckle \\ \hline
& Shane (3 m) & ShARCS & Ks & AO \\ \hline
1616 & & & & \\ \hline
& Shane (3 m) & ShARCS & Ks & AO \\ \hline
& Shane (3 m) & ShARCS & J & AO \\ \hline
& \textbf{SAI-2.5m (2.5 m)} & \textbf{Speckle Polarimeter} & \textbf{I} & \textbf{Speckle} \\ \hline
2719 & & & & \\ \hline
& SOAR (4.1 m) & HRCam & I & Speckle \\ \hline
& \textbf{Palomar (5 m)} & \textbf{PHARO} & \textbf{Kcont} & \textbf{AO} \\ \hline
4051 & & & & \\ \hline
& \textbf{SAI-2.5m (2.5 m)} & \textbf{Speckle Polarimeter} & \textbf{I} & \textbf{Speckle} \\ \hline
4094 & & & & \\ \hline
& \textbf{SAI-2.5m (2.5 m)} & \textbf{Speckle Polarimeter} & \textbf{I} & \textbf{Speckle} \\ \hline
& Shane (3 m) & ShARCS & Ks & AO \\ \hline
& Shane (3 m) & ShARCS & J & AO \\ \hline
4155 & & & & \\ \hline
& WIYN (3.5 m) & NESSI & 562 (44) nm & Speckle \\ \hline
& SAI-2.5m (2.5 m) & Speckle Polarimeter & I & Speckle \\ \hline
& \textbf{WIYN (3.5 m)} & \textbf{NESSI} & \textbf{832 (40) nm} & \textbf{Speckle} \\ \hline
4731 & & & & \\ \hline
& \textbf{SAI-2.5m (2.5 m)} & \textbf{Speckle Polarimeter} & \textbf{I} & \textbf{Speckle} \\ \hline
5706 & & & & \\ \hline
& Palomar (5 m) & PHARO & Kcont & AO \\ \hline
& WIYN (3.5 m) & NESSI & 832 (40) nm & Speckle \\ \hline
& Palomar (5 m) & PHARO & Hcont & AO \\ \hline
& SAI-2.5m (2.5 m) & Speckle Polarimeter & I & Speckle \\ \hline
& \textbf{Gemini (8 m)} & \textbf{'Alopeke} & \textbf{832 (40) nm} & \textbf{Speckle} \\ \hline
& Gemini (8 m) & 'Alopeke & 562 (54) nm & Speckle \\ \hline
& WIYN (3.5 m) & NESSI & 562 (44) nm & Speckle \\ \hline
5735 & & & & \\ \hline
& \textbf{Keck2 (10 m)} & \textbf{NIRC2} & \textbf{Kcont} & \textbf{AO} \\ \hline
6000 & & & & \\ \hline
& \textbf{SAI-2.5m (2.5 m)} & \textbf{Speckle Polarimeter} & \textbf{I} & \textbf{Speckle} \\ \hline
6324 & & & & \\ \hline
& \textbf{Keck2 (10 m)} & \textbf{NIRC2} & \textbf{Kcont} & \textbf{AO} \\ \hline
6397 & & & & \\ \hline
& \textbf{SAI-2.5m (2.5 m)} & \textbf{Speckle Polarimeter} & \textbf{I} & \textbf{Speckle} \\ \hline
\end{tabular}
\caption{Details of high-resolution imaging. The contrast curves used in \texttt{TRICERATOPS} are in bold.}
\label{tab:imaging}
\end{table*}

\newpage
\begin{figure*}
    \centering
    \includegraphics[width=0.99\textwidth]{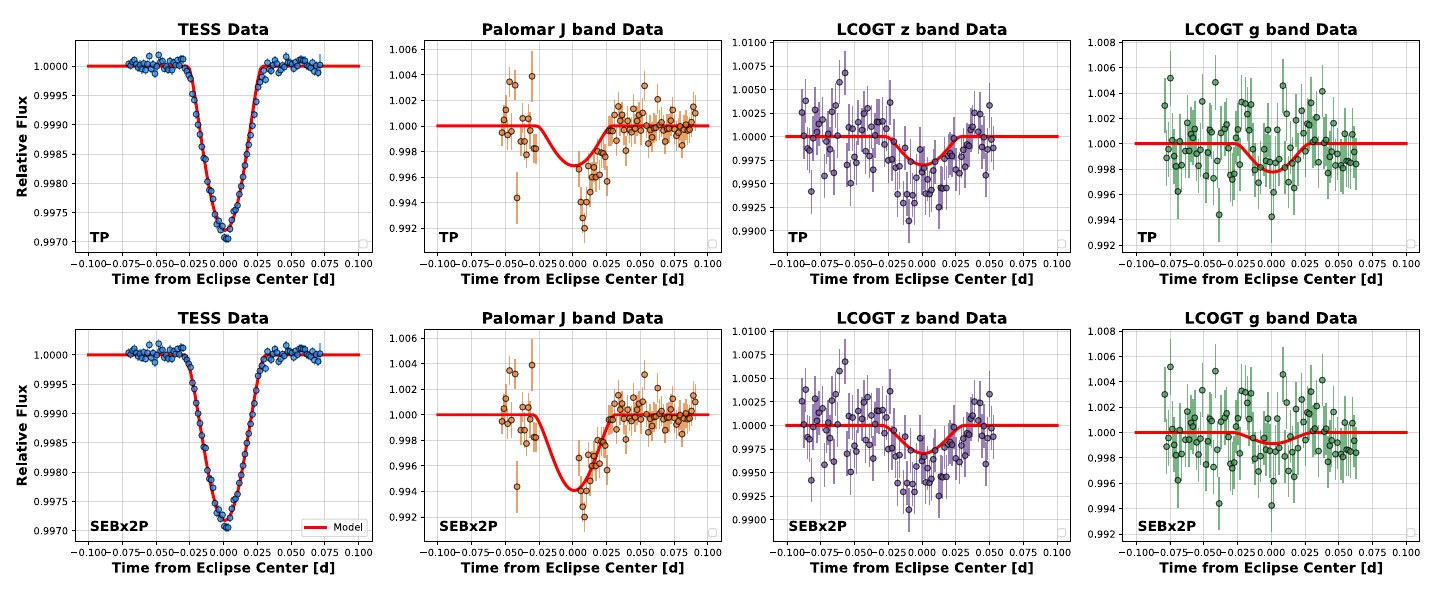}
    \caption{\textit{Top:} TRICERATOPS+ fit (red curve) for the TP scenario (transiting planet around target star) to the observations of TOI-1254.01. \textit{Bottom:} TRICERATOPS+ fit (red curve) for the SEBx2P scenario (an unresolved eclipsing binary with twice the orbital period around a secondary star) compared to the same observational dataset. This scenario has a relative probability of 0.9999.}
    \label{fig:tricer_fits}
\end{figure*}

\bibliography{tess_palomar}{}
\bibliographystyle{aasjournal}

\end{document}